# Temperature Distributions in the Protosatellite Disks of Uranus, Jupiter, Neptune and Saturn and Common Structure in Their Satellite Systems

James C. Lombardi Sr. Professor Emeritus,
Physics Department, Allegheny College, Meadville, PA, USA; james.lombardi@allegheny.edu

ABSTRACT

We observe similar structures in the orbital radii of satellite and ring systems of Uranus, Jupiter and Neptune. This stimulates an investigation that the evolution of these systems and Saturn's system is initiated by a common mechanism that involves the interaction of radiation with their subnebulae (protosatellite disks). A model is presented that is characterized by resonance created through stimulated radiative molecular association (SRMA) reactions. In this model thermal energy is extracted from a protoplanetary disk at specific distances from the protoplanet wherever there is a match between the local thermal energy of the disk and the energy of photons impinging on the disk. Radiation is supplied by a portion of the hydrogen molecule's spectrum for the Uranian, Jovian and Neptunian disks and a portion of the hydrogen atom's spectrum for the Saturnian disk. Findings shed light on the early evolution and structure of satellite systems including the complicated ring system of Saturn. They also link satellite orbital radii size distributions to shapes of temperature distributions (TD's) in protosatellite disks. Theoretically determined TD's in the protosatellite disks of Uranus (Mousis 2004) and Saturn (Mousis Gautier and Bockelee-Moran, 2002) are essential to the present investigation.

## 1. INTRODUCTION

This investigation deals with the satellites that are created in the mid-planes of the protosatellite disks of Uranus, Jupiter, Neptune and Saturn. These satellites have orbits with small eccentricities and inclinations relative to the planet's equatorial plane and are generally called regular satellites. However for ease of discussion here they are usually simply referred to as satellites. Section 2.1 introduces relationships among the orbital radii of regular satellites and rings of Uranus, Jupiter and Neptune. Subsequent sections present a model that explains these relationships and also the structure of the satellite systems of the four giant gas planets in the solar system. In this model regular satellites are formed where disk temperatures have certain values with these values depending on the energies of photons that exist in the disk. We propose the connection between temperature and photon energy to be a result of resonance involving stimulated radiative molecular association (SRMA).

In this investigation TD stands for the midplane temperature distribution or portion of a temperature distribution in the protosatellite disk of a protoplanet. Results of this investigation rely heavily on the theoretically determined TD's in the protosatellite disks of Uranus (Mousis 2004) and Saturn (Mousis Gautier and Bockelee-Moran, 2002).

This paper is a rewrite of Lombardi (2015a). Lombardi (2015b) was a follow-up for Lombardi (2015a) and now it should be a follow-up for this paper.

## 2. RESULTS AND DISCUSSION

Table 1 contains the names of known regular satellites of Uranus, Jupiter, and Neptune and their orbital radii (lengths of the semi-major axis) in units of $R_U = 25{,}559$ km, $R_J = 71{,}492$ km and $R_N = 24{,}766$ km, the equatorial radii of Uranus, Jupiter and Neptune respectively (NASA 2021). I.e. if $r$ is the orbital radius of a Uranian satellite in km, then $r/R_U$ is its orbital radius in units of $R_U$. In the case of Uranus, satellites with orbital radii less than $r/R_U = 2.3$ are not listed in Table 1 but are considered later. For Jupiter and Neptune all of the known regular satellites are listed. Values for the index $i$ are listed in the first column of Table 1. It will become apparent why the indices have their particular listed values in later sections. All satellites and orbital radii in the same row are associated with the same value of $i$. Note that Cupid in the Uranian system does not have a value of $i$ assigned to it. The reason for this is in subsection *2.1.b*.

*2.1.a. The Linearity of the Uranian vs Jovian and
Uranian vs Neptunian Orbital Radii*

Fig. 1 is a graph of orbital radii of a certain set of contiguous Uranian satellites (excluding Cupid) versus the orbital radii of all the Jovian satellites. Each point is for the two radii with the same $i$ value. No other set of pairings of Uranian and Jovian satellite orbital radii produces a graph that is fitted by a straight line nearly as well as the one used to make Fig. 1. In Figs. 1 & 2 and Eqs. (1) & (2), $R_{ui} = r/R_U$ for the $i^{th}$ Uranian satellite orbital radius. Similar definitions hold for $R_{ji}$ and $R_{ni}$.

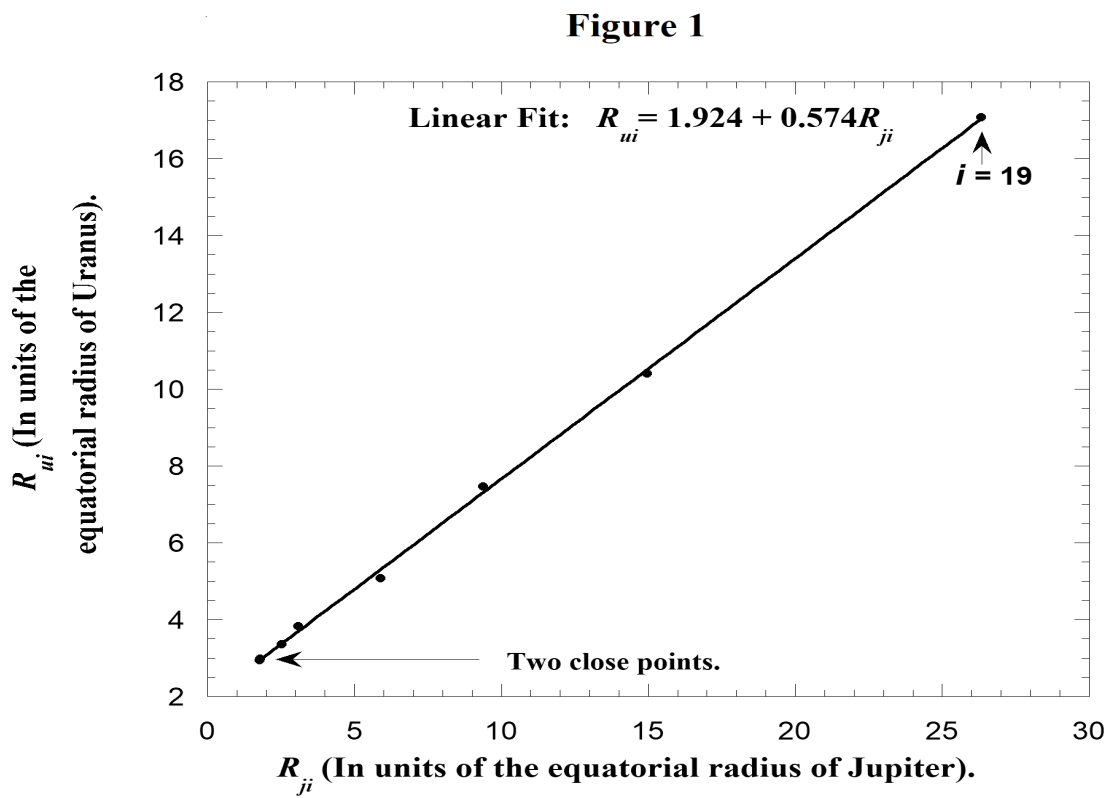

**Fig. 1.** All the Jovian satellite orbital radii are linearly related to a certain set of Uranian satellite orbital radii.

**Figure 1**

Linear Fit: $R_{ui} = 1.924 + 0.574 R_{ji}$

$i = 19$

Two close points.

$R_{ui}$ (In units of the equatorial radius of Uranus).

$R_{ji}$ (In units of the equatorial radius of Jupiter).

Fig. 2 is a graph that is similar to Fig. 1 except Fig. 2 is for Uranian and Neptunian satellites. The graph is also well fitted by a straight line except for the Mab-Hippocamp point where $i = 15$ as seen in Table 1. This discrepancy is discussed in subsection 2.1.b.

**Fig. 2. All but one of the Neptunian satellite orbital radii are linearly related to a certain set of Uranian satellite orbital radii.**

**Figure 2**

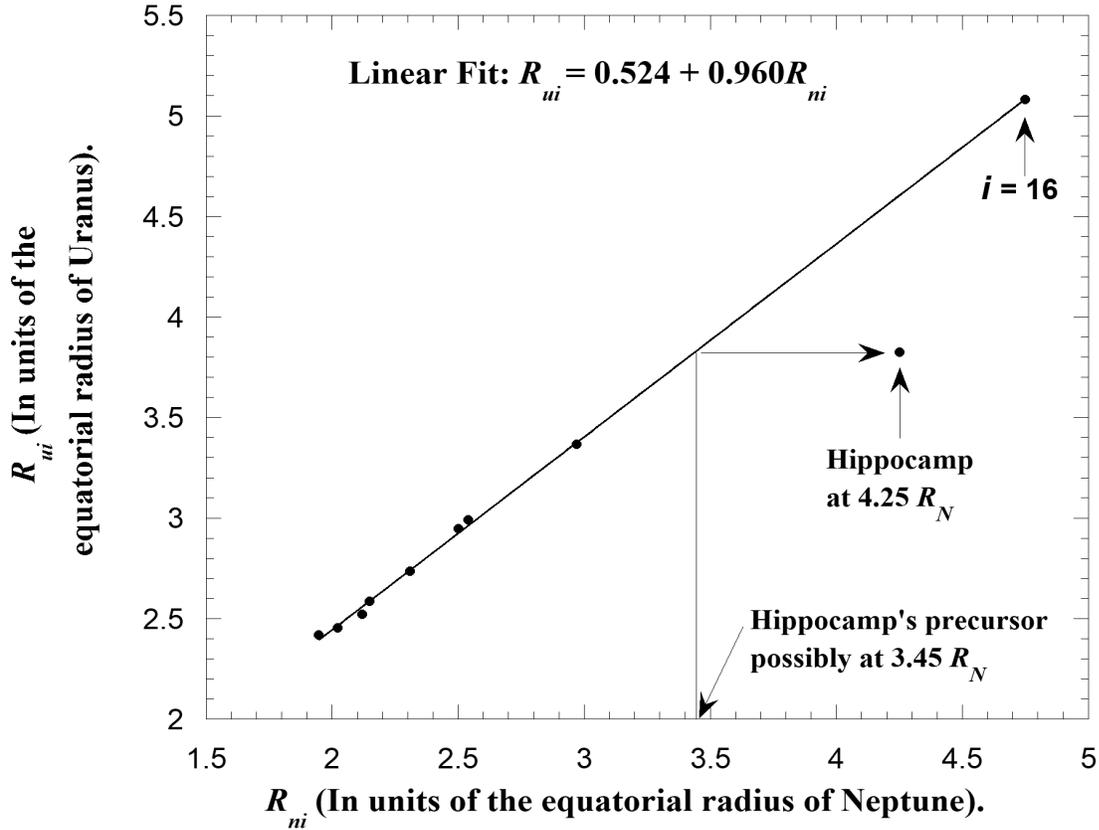

The equations of the best-fit lines to the graphs in Figs. 1 and 2 are

$$R_{ui} = 1.924 + 0.574 R_{ji} \quad (1)$$

and

$$R_{ui} = 0.524 + 0.960 R_{ni}, \quad (2)$$

Ring Galle in Neptune's system lies inside the innermost satellite. Its width is large and optical depth low (NASA 2021). Ring Galle is not included in the present analysis.

*2.1.b. Details Concerning Cupid and Hippocamp*

We mentioned above that Cupid in the Uranian system does not have a value of $i$ assigned to it in Table 1 nor is it included in future analysis. Cupid, Belinda $(i = 11)$ and Perdita $(i = 12)$ are part of a triplet in which each member has nearly the same orbital radius. This triplet is associated with doublets in the system of Jovian and Neptunian systems. If we assign a value of $i$ to each of the three satellites in the triplet, we do not obtain good linear fits as seen in Figs. 1 and 2. Possibly sometime during the evolution of a doublet in the Uranian system the two members collide resulting in the triplet as seen in Table 1.

The Mab-Hippocamp point (where $i$ =15 for both satellites) in Fig. 2 is far from the linear fit. Mab's orbital radius together with Thebe's orbital radius creates a point in Fig. 1. The fact that the points in Fig. 1 (including Mab's) are well fitted by a straight line suggests strongly that, the Mab-Hippocamp point in Fig. 2 is not positioned well because of the value of Hippocamp's orbital radius, not Mab's. Table 2 indicates that Hippocamp's mass is small compared to the masses of the other nearby satellites. Perhaps a precursor of Hippocamp experiences an impact that shatters it leaving Hippocamp as a fragment. Indicated in Fig. 2 are the orbital radius of $4.25R_N$ for Hippocamp and the approximate orbital radius of $3.45R_N$ for its precursor.

### 2.1.c. Eqs. (1) & (2) as Transformation Equations for Orbital Radii

Consider Table 1. The Jovian distribution of orbital radii starts with $R_{j11}$ and ends with $R_{j19}$. Eqs. (1) can be used to scale and shift the radial coordinates that describe the Jovian system so all the $R_{ji}$ from $R_{j11}$ to $R_{j19}$ span the same normalized radial coordinates as does the portion of the Uranian system from $R_{u11}$ to $R_{u19}$. The $0.574R_{ji}$ term in Eq. (1) scales the radial distance associated with the $R_{ji}$ and the constant 1.924 shifts the origin of the system. Therefore if we substitute a value of $R_{ji}$ into Eq. (1) we calculate a transformed value of $R_{ji}$ (call it $R_{Tji}$) that is close to the corresponding value of $R_{ui}$. The same idea holds for the transformation of the $R_{ni}$ values (call each one $R_{Tni}$) by using Eq. (2). Table 3 contains the scaled (transformed) orbital radii of satellites of Jupiter and Neptune along with the orbital radii of all the regular satellites and rings of Uranus (NASA 2021).

To visualize the usefulness of transformed orbital radii, consider Fig. 3 which is constructed using Table 3. (The particular $i$ values listed in Table 3 will be discussed later). The top panel is the distribution of transformed Neptunian orbital radii. The middle panel is the distribution of transformed Jovian orbital radii. The bottom panel is the distribution of all the Uranian satellite orbital radii, not just those that overlap with the Neptunian and Jovian satellites. Notice the Neptunian and Jovian line graphs have points (actually vertical lines or large dots) that line up closely with radii in the Uranian distribution and also with each other. The single exception is Hippocamp's point as expected.

The positions corresponding to the surfaces of Jupiter and Neptune are each determined by substituting the number 1 into the right side of Eq. (1) and Eq. (2) respectively. The transformed surface radial coordinates for the Jovian disk and the Neptunian disk are 2.498 and 1.484 respectively. These are both compared to the Uranian surface coordinate of 1. The fact that they are larger than 1 indicates that the region where the Uranian satellites are created is larger than the corresponding normalized regions for Jupiter and Neptune. This may help to explain why there are more satellites in the Uranian system than in each of the Jovian and Neptunian systems.

Table 1. Orbital radii of satellites and rings of Uranus, Jupiter and Neptune in units of their equatorial radii. For instance if $R_U$ is the equatorial radius of Uranus in km and $r_{ui}$ is the orbital radius of a Uranian satellite in km, then $R_{ui} = r_{ui}/R_U$ is its orbital radius in units of $R_U$.

| $i$ | Uranian Satellites | $R_{ui}$[a] | Jovian Satellites | $R_{ji}$[a] | Neptunian Satellites | $R_{ni}$[a] |
|---|---|---|---|---|---|---|
| 1 | Bianca | 2.316 | | | | |
| 3&4 | Cressida | 2.418 | | | Naiad | 1.947 |
| 5&6 | Desdemona | 2.453 | | | Thalassa | 2.022 |
| 7 | Juliet | 2.520 | | | Despina | 2.121 |
| 9 | Portia | 2.586 | | | Rings LeV&Las[c] | 2.148 |
| 10 | Rosalind | 2.735 | | | Ring Arago | 2.310 |
| | Cupid[b] | 2.911 | | | | |
| 11 | Belinda | 2.946 | Metis | 1.790 | Galatea & Unnamed ring | 2.502 |
| 12 | Perdita | 2.990 | Adrastea | 1.804 | Ring Adams | 2.541 |
| 13&14 | Puck | 3.365 | Amalthea | 2.537 | Larissa | 2.970 |
| 15 | Mab | 3.824 | Thebe | 3.104 | Hippocamp | 4.252 |
| 16 | Miranda | 5.082 | Io | 5.900 | Proteus | 4.750 |
| 17 | Ariel | 7.469 | Europa | 9.387 | | |
| 18 | Umbriel | 10.407 | Ganymede | 14.972 | | |
| 19 | Titania | 17.070 | Callisto | 26.334 | | |
| 20 | Oberon | 22.830 | | | | |

Note: Each satellite is assigned an index ($i$) consisting of one or two integers. Satellites in the same row have the same $i$. The indexing system is explained in section 2.2.
[a]NASA(2021)
[b]Section 2.1.b explains why Cupid does not have an index nor is it used in the analysis.
[c]Rings LeV&Las stands for Rings LeVerrier and Lassell

Table 2. Orbital radii and masses of satellites of Neptune. The list includes all satellites in the group nearest to Neptune most of which are regular satellites. Rings are not included.

| $i$ | | | Orbital Radius $R_{ni}$[a] | Mass[a] ($10^{17}$ kg) |
|---|---|---|---|---|
| 3&4 | | Naiad | 1.948 | 2 |
| 5&6 | | Thalassa | 2.022 | 4 |
| 7 | | Despina | 2.121 | 20 |
| 11 | | Galatea | 2.502 | 40 |
| 13&14 | | Larissa | 2.970 | 50 |
| 15 | | Hippocamp | 4.252 | 0.3 |
| 16 | | Proteus | 4.751 | 500 |
| — | | Triton[b] | 14.328 | 214000 |
| — | | Nereid[b] | 222.67 | 300 |

[a]NASA(2021)
[b]Triton and Nereid are irregular satellites. Their orbital plane is not aligned with the orbital plane of the regular satellites.

Table 3. Orbital radii of satellites and rings of Uranus, Jupiter and Neptune in units of equatorial radius of the respective planet. For Jupiter and Neptune, $R_{Tji}$ and $R_{Tni}$ are orbital radii transformed using Eqns (1) and (2) respectively.

| $i$ | Uranian Satellites | $R''_{ui}$ [a] | $R'_{ui}$ [a] | $R_{ui}$ [a] | Jovian Satellites | $R_{Tji}$ [b] | Neptunian Satellites | $R_{Tji}$ [c] |
|---|---|---|---|---|---|---|---|---|
| 11&12 | Ring 6 | 1.637 | | | | | | |
| 13&14 | Ring 5 | 1.652 | | | | | | |
| 15 | Ring 4 | 1.666 | | | | | | |
| 15 | Ring alpha | | 1.750 | | | | | |
| 13&14 | Ring β | | 1.787 | | | | | |
| 11&12 | Ring η | | 1.846 | | | | | |
| 10 | Ring γ | | 1.863 | | | | | |
| 9 | Ring δ | | 1.900 | | | | | |
| 7 | Cordelia | | 1.948 | | | | | |
| 5&6 | Ring λ | | 1.957 | | | | | |
| 3&4 | Ring ε | | 2.006 | | | | | |
| 1 | Ophelia | | 2.105 | | | | | |
| 1 | Bianca | | | 2.316 | | | | |
| 3&4 | Cressida | | | 2.418 | | | Naiad | 2.393 |
| 5&6 | Desdemona | | | 2.453 | | | Thalassa | 2.465 |
| 7 | Juliet | | | 2.520 | | | Despina | 2.560 |
| 9 | Portia | | | 2.586 | | | Rings LeV&Las [d] | 2.586 |
| 10 | Rosalind | | | 2.735 | | | Ring Arago | 2.742 |
| | Cupid [e] | | | 2.911 | | | | |
| 11 | Belinda | | | 2.946 | Metis | 2.951 | Galatea & Unnamed ring | 2.926 |
| 12 | Perdita | | | 2.990 | Adrastea | 2.959 | Ring Adams | 2.963 |
| 13&14 | Puck | | | 3.365 | Amalthea | 3.380 | Larissa | 3.375 |
| 15 | Mab | | | 3.824 | Thebe | 3.706 | Hippocamp | 4.606 |
| 16 | Miranda | | | 5.082 | Io | 5.311 | Proteus | 5.084 |
| 17 | Ariel | | | 7.469 | Europa | 7.312 | | |
| 18 | Umbriel | | | 10.407 | Ganymede | 10.518 | | |
| 19 | Titania | | | 17.070 | Callisto | 17.040 | | |
| 20 | Oberon | | | 22.830 | | | | |

Note: Each satellite is assigned an index ($i$) consisting of one or two integers. Satellites in the same row have the same $i$. The indexing system is explained in sections 2.1.c, 2.2a and 2.2b.

[a] NASA(2021)

$R''_{ui}$ refers to orbital radii from Ring 6 to Ring 4.

$R'_{ui}$ refers to orbital radii from Ring α to Ophelia.

$R_{ui}$ refers to orbital radii from Bianca to Oberon.

[b] NASA(2021) transformed with Eqn (1) as described in sections 2.1.c

[c] NASA(2021) transformed with Eqn (2) as described in sections 2.1.c

[d] Rings LeV&Las stands for Rings LeVerrier and Lassell

[e] See text concerning why the orbital radius of Cupid does not have an index nor is it used in the analysis.

**Fig. 3. Orbital radii of the Jovian and Neptunian satellites line up with the orbital radii of the Uranian satellites after Eqs. (1) and (2) are used to transform the Jovian and Neptunian orbital radii.**

Figure 3

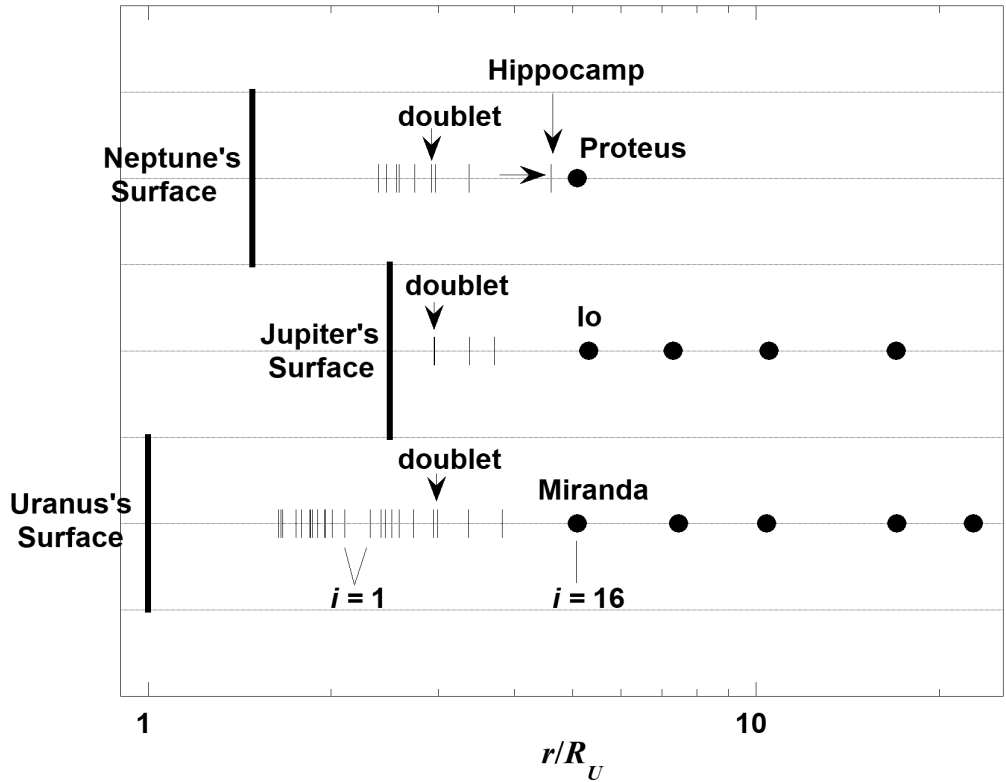

In Fig. 3 the large dots in the Uranian line graph correspond to Uranus's major satellites. In the Jovian graph large dots represent the Galilean satellites. In the Neptunian graph Proteus is by far the most massive regular satellite as seen in Table 3. We observe the most massive satellites exist at and beyond orbital radii for which $i = 16$. Later discussions in the present investigation will link distributions of satellite orbital radii to protosatellite disks. Fig. 3 and this linkage indicates that massive satellites are created in similar regions of protosatellite disks.

*2.2.a. A Smooth Relationship Exists Between Photon Energies in the*
*$H_2$ Spectrum and the Orbital Radii of Uranian Satellites*

The linearity of the graphs in Figs. 1 and 2 suggests a common mechanism creates the primordial rings around Uranus, Jupiter and Neptune which evolve into satellites. The present investigation explores the possibility that photons corresponding to molecular hydrogen ($H_2$) lines interact with molecules in the protosatellite disks of Uranus, Jupiter and Neptune and through the SRMA mechanism they initiate resonance that cools specific rings (resonance rings) that encircle the protoplanets. SRMA and its associated resonance is discussed in section 2.6. In preparation for section 2.6, this and the next few sections discuss the relationship that exists between photon energies and the orbital radii of Uranian satellites.

$H_2$ lines are prominent in the spectra of nebulae. Draine and Bertoldi (1996) have modeled the spectrum of the reflection nebula NGC 2023, and Martini et al. (1999) have observed the spectra of four reflection nebulae including NGC 2023. In both of these investigations, the large spectral peaks in the range 5560 – 4290 cm$^{-1}$ (1.80 – 2.33 μm) all correspond to S-branch transitions (lines) in the $H_2$ spectrum. The segment of the $H_2$ spectrum that we focus on is shown in Fig. 1 of Martini et al. (1999) which is a model spectrum for the reflection nebulae they observed. Fig. 4 in the present investigation is a similar model spectrum developed using photon energies ($E_p$'s) and spectral intensities ($I$'s) from Black and van Dishoeck (1987). $E_p$'s in Fig. 4 and throughout this paper are in wave numbers (cm$^{-1}$). Also in Fig. 4 each index $i$ is associated with a particular S-branch transition. That is, each $i$ value stands for a particular transition in $H_2$.

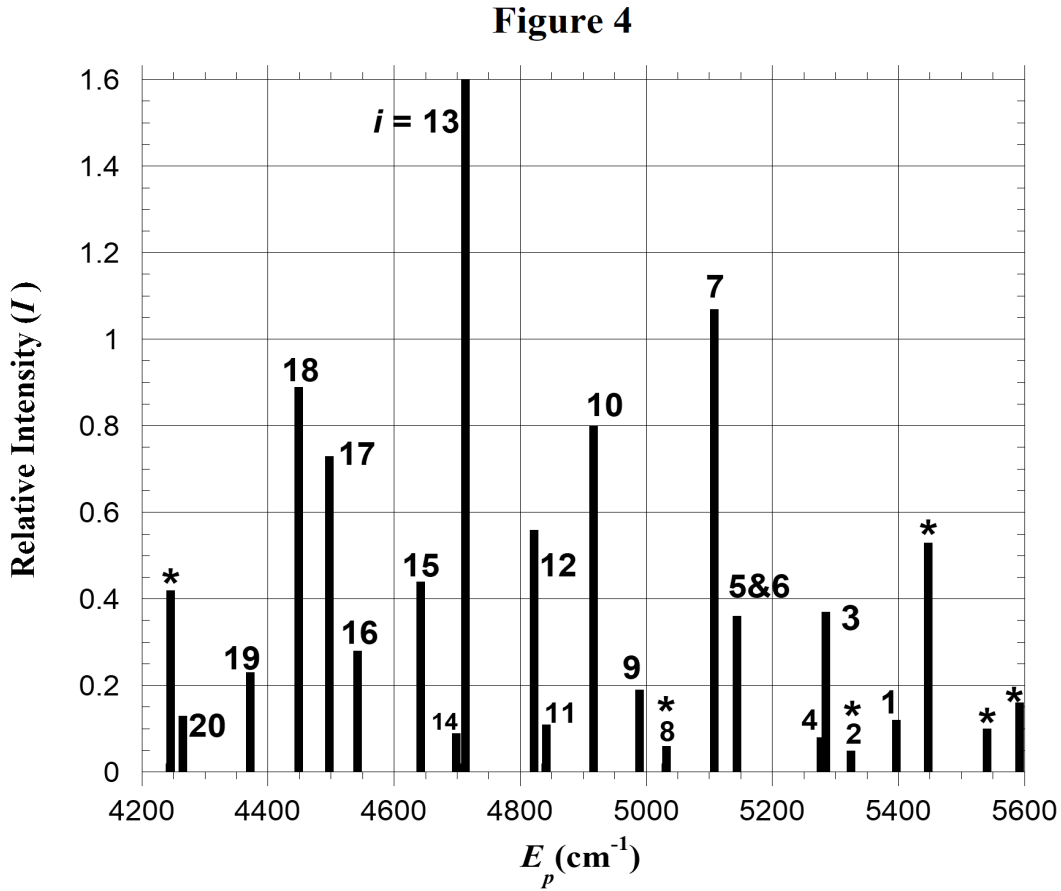

Fig. 4. $H_2$ S-Branch Spectrum: Starred(*) spectral lines fall outside the range of peaks that affect the Uranian disk or correspond to intensities that are too low to cause resonance.

**Figure 4**

The origin of the photons in the $H_2$ spectrum that initiate the key SRMA reactions in a disk is not identified in the present investigation. Perhaps they come from $H_2$ molecules that reside in the the disk. Or perhaps they come from the protosun, presumably a typical T Tauri star. The $H_2$ spectrum is observed as a component of the overall spectra of T Tauri stars (Beckwith, Gatley, Matthews and Neugebauer 1978, and Herczeg et al. 2006). Furthermore, FUors and EXors (Hartmann, Kenyon and Hartigan 1993) are subclasses of T Tauri's. These objects are characterized by short-lived magnitude changes of up to 6 magnitudes with rise times on the order of months to years (Appenzeller and Mundt 1989, and Herbig 1977). Possibly the protosun in such a short lived state delivers the photons that initiate resonance rings in protoplanetary disks.

Table 4 lists the complete set of regular Uranian satellites and their orbital radii ($R_{ui}$) in units of the equatorial radius of Uranus. The subscript *i* is a reminder that each orbital radius is associated with an $E_p$ value or in some cases a weighted sum of two $E_p$'s. A graph of $E_p$ vs $R_{ui}$ is referred to as a photon energy distribution (PED). The PED in Fig. 5a is constructed from the data in Table 4. Consider the main features of the PED, i.e. the dip, peak and the long tail. The method used for establishing the particular set of pairings ($R_{ui}$, $E_p$) in Table 4 is given in subsection 2.2.b. Later, more credence is given to the concept of the PED when it is linked to the temperature distribution (TD) of the Uranian protosatellite disk.

Fig. 5a. The complete PED in the Uranian protoplanetary disk

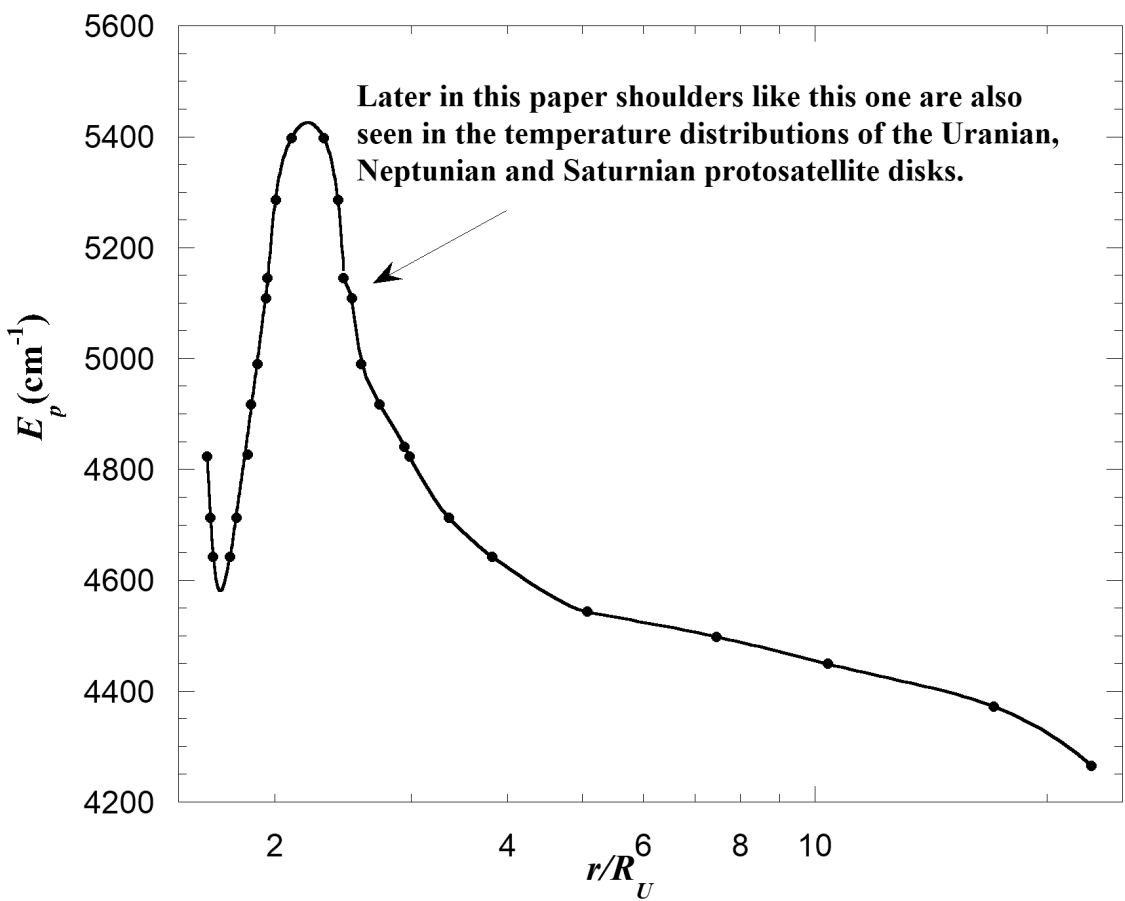

Table 4. $H_2$ Transitions and their associated photon energies ($E_P$'s) and relative spectral intensities ($I$'s) associated with Uranian satellite orbital radii ($R''_{ui}$, $R'_{ui}$ and $R_{ui}$)

| $i$ | $H_2$ Transition[a] | $E_P$(cm⁻¹)[a] | $I$[a] | $E_P$'s for Figs. 5a, 5b and 5c | Uranian Satellite | ($R''_{ui}$ or $R'_{ui}$)[b] |
|---|---|---|---|---|---|---|
| **11** | **(3,2) S(5)** | **4841** | **0.11** | | | |
| | | | | } 4826[c] | **Ring 6** | **1.637** |
| **12** | **(2,1) S(3)** | **4823** | **0.56** | | | |
| **13** | **(1,0) S(1)** | **4713** | **1.6** | | | |
| | | | | } 4712[c] | **Ring 5** | **1.652** |
| **14** | **(3,2) S(4)** | **4699** | **0.09** | | | |
| 15 | (2,1) S(2) | 4642 | 0.44 | 4642 | Ring 4 | 1.666 |
| 15 | (2,1) S(2) | 4642 | 0.44 | 4642 | Ring α | 1.750 |
| 14 | (3,2) S(4) | 4699 | 0.09 | | | |
| | | | | } 4712[c] | **Ring β** | **1.786** |
| 13 | (1,0) S(1) | 4713 | 1.6 | | | |
| 12 | (2,1) S(3) | 4823 | 0.56 | | | |
| | | | | } 4826[c] | **Ring η** | **1.834** |
| 11 | (3,2) S(5) | 4841 | 0.11 | | | |
| 10 | (1,0) S(2) | 4917 | 0.8 | 4917 | Ring γ | 1.863 |
| 9 | (2,1) S(4) | 4990 | 0.19 | 4990 | Ring δ | 1.900 |
| 8 | (9,7) S(0)[d] | 5032 | 0.06 | | | |
| 7 | (1,0) S(3) | 5108 | 1.07 | 5108 | Cordelia | 1.948 |
| 6 | (2,1) S(5) | 5142 | 0.25 | | | |
| | | | | } 5144[c] | **Ring λ** | **1.957** |
| 5 | (9,7) S(1) | 5147 | 0.11 | | | |
| 4 | (2,1) S(6) | 5278 | 0.08 | | | |
| | | | | } 5285[c] | **Ring ε** | **2.006** |
| **3** | **(1,0) S(4)** | **5286** | **0.37** | | | |
| 2 | (9,7) S(3)[d] | 5325 | 0.05 | | | |
| 1 | (2,1) S(7) | 5397 | 0.12 | 5397 | Ophelia | 2.105 |

Table 4 continued

| $i$ | H$_2$ Transition[a] | $E_P$(cm$^{-1}$)[a] | $I$ [a] | $E_P$'s for Fig 5 | Uranian Satellite | $R_{ui}$ [b] |
|---|---|---|---|---|---|---|
| 1 | (2,1) S(7) | 5397 | 0.12 | 5397 | Bianca | 2.316 |
| 2 | (9,7) S(3)[d] | 5325 | 0.05 | | | |
| **3** | **(1,0) S(4)** | **5286** | **0.37** | | | |
| | | | | } 5285[c] | **Cressida** | **2.418** |
| **4** | **(2,1) S(6)** | **5278** | **0.08** | | | |
| **5** | **(9,7) S(1)** | **5147** | **0.11** | | | |
| | | | | } 5144[c] | **Desdemona** | **2.453** |
| **6** | **(2,1) S(5)** | **5142** | **0.25** | | | |
| 7 | (1,0) S(3) | 5108 | 1.07 | 5108 | Juliet | 2.520 |
| 8 | (9,7) S(0)[d] | 5032 | 0.06 | | | |
| 9 | (2,1) S(4) | 4990 | 0.19 | 4990 | Portia | 2.586 |
| 10 | (1,0) S(2) | 4917 | 0.8 | 4917 | Rosalind | 2.735 |
| 11 | (3,2) S(5) | 4841 | 0.11 | 4841 | Belinda | 2.946 |
| 12 | (2,1) S(3) | 4823 | 0.56 | 4823 | Perdita | 2.990 |
| **13** | **(1,0) S(1)** | **4713** | **1.6** | | | |
| | | | | } 4712[c] | **Puck** | **3.365** |
| **14** | **(3,2) S(4)** | **4699** | **0.09** | | | |
| 15 | (2,1) S(2) | 4642 | 0.44 | 4642 | Mab | 3.824 |
| 16 | (3,2) S(3) | 4543 | 0.28 | 4543 | Miranda | 5.082 |
| 17 | (1,0) S(0) | 4498 | 0.73 | 4498 | Ariel | 7.469 |
| 18 | (2,1) S(1) | 4449 | 0.89 | 4449 | Umbriel | 10.407 |
| 19 | (3,2) S(2) | 4372 | 0.23 | 4372 | Titania | 17.070 |
| 20 | (4,3) S(3) | 4265 | 0.13 | 4265 | Oberon | 22.830 |

[a]Black van Dishoeck (1987)

[b]NASA (2021)

$R''_{ui}$ refers to orbital radii from Ring 6 to Ring 4.

$R'_{ui}$ refers to orbital radii from Ring α to Ophelia.

$R_{ui}$ refers to orbital radii from Bianca to Oberon.

[c]A bolded $E_P$ is a weighted average of the $E_P$'s on the previous and

following lines. The weighting factors are the corresponding $I$'s.

[d]It is not possible to associate this low intensity spectral line with any satellite.

Note that spectral lines appear in Table 4 one to three times because the PED has a dip and a peak in it. Also Fig. 4 shows closely spaced lines corresponding to indices 3&4, 5&6 and 13&14. In these cases a weighted average of $E_p$'s is used with the weighting factors being the corresponding spectral intensities. The closely spaced lines with indices 11&12 have their $E_p$'s averaged for ring 6 and ring η (ring η is known to have two components (French et al. 1991)). But they are resolved for the "doublet" Belinda and Perdita. Two spectral lines with indices 2 and 8 are omitted from the analysis because they have low relative intensity as is indicated in Table 4 and Fig. 4 and they cannot be paired with orbital radii. Cupid's orbital radius is also omitted from the analysis as previously mentioned.

*2.2.b. The Determination of the Relationship Between $E_p$ Values and*

*Orbital Radii Given in Table 4 and Figs. 5a & 5b*

The following are the steps taken to develop the PED's in Figs. 5a and 5b. By trial and error it is determined that if the $R_{ui}$'s for rings α, β, η, γ, δ and satellite Cordelia are paired with $E_p$'s corresponding to $i$'s 15, 14&13, 12&11, 10, 9 and 7 respectively as in Table 4, then a plot of $E_p$ vs $R_{ui}$ is well fitted by a straight line shown by the straight dashed portion of the curve in Fig. 5b. The point for ring η is noticeably the furthest away from the straight line fit. It is interesting that the radius of ring η is very near the 3:2 Inner Lindblad Resonance of the satellite Cressida (Chancia et al. 2017). Perhaps the radius of ring η has been affected by its interaction with Cressida.

Fig. 5b. The photon energy distribution from Ring 6 to Desdemona

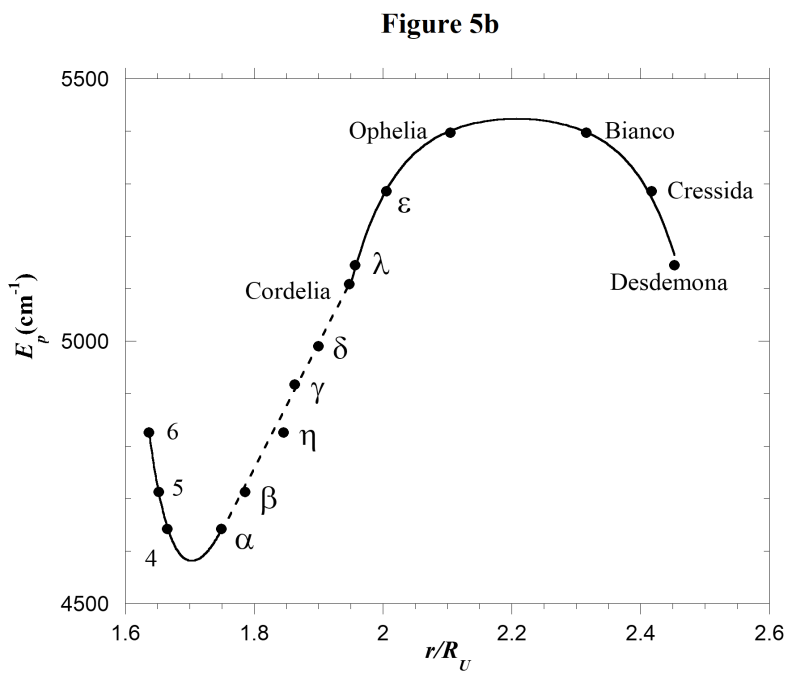

**Figure 5b**

When we pair rings 6, 5 and 4 (the three innermost rings) incorrectly with $i$'s 18, 17 and 16 (which might be the expected inward extension of $i$'s) a discontinuity is created as seen in Fig. 5c. If instead we use the $i$'s 12&11, 14&13 and 15 for the three innermost rings a smoothly varying dip becomes apparent as seen in Fig. 5b.

**Fig. 5c. The Photon Energy Distribution with Discontinuity**

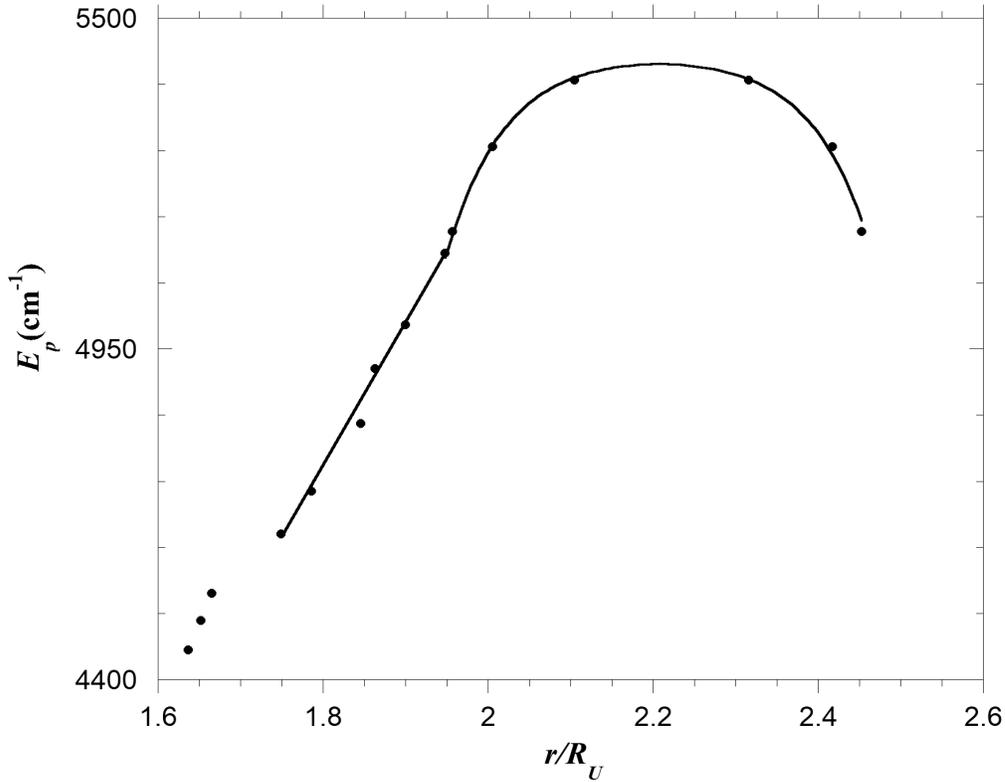

**Figure 5c**

A similar situation occurs for the region from Cordelia to Desdemona in Fig. 5b. If we don't assume a peak occurs there, a discontinuity occurs because of the wide radial gap between Ophelia and Bianca. If we assume a peak, then we find the curve is smoothly varying, symmetric and centered at $r/R_U =$ 2.21. This point is half-way between Ophelia and Bianca as seen in Fig. 5b.

Notice that Bianca is assigned $i = 1$ just as for Ophelia. In Table 4 and Fig. 5a, the rest of the PED beyond Bianco does not have any more dips or peaks and each satellite in turn is assigned an ever increasing $i$ or pair of $i$ 's all the way out to $i = 20$ for Oberon the last regular satellite.

*2.3.a. Comparing the PED in Figure 5a to TD's of the Solar Nebula*

We now compare the PED in Figs. 5a to temperature distributions (TD's) calculated by Lin and Papaloizou (1985, hereafter L&P (1985)) for the solar nebula. Fig.18 of L&P (1985) contains two sets of graphs showing midplane TD's of the solar nebula for two different times during what are called upward and downward transition waves. We find the PED in Fig. 5b to be similar in shape to the L&P (1985) solar nebula TD's. Both types of L&P (1985) TD's are peaked and have characteristics similar to those of the PED in Fig. 5a. To the right of the peak in the L&P (1985) TD at the onset of the transition, there is a negatively sloped tail just as in the PED in Fig. 5a. Also, in the L&P (1985) distribution near the end of the downward transition and to the left of the peak, the TD has a dip just as in the PED in Fig. 5a.

### 2.3.b. The PED and TD of the Uranian Protosatellite Disk

The observations in the last subsection cause us to explore the possibility that the PED in Fig. 5a is related to the TD of the Uranian protosatellite disk. To do this we first make a series of simplifying assumptions: 1. The present day orbital radii of the Uranian satellites and rings ($R_u$) are the same or nearly the same as the orbital radii of the primordial rings from which they evolved. If migration does affect Uranian satellite orbital radii, the migration is uniform for all the satellites. 2. The Uranian rings and regular satellites evolved from primordial rings that were born in the Uranian protosatellite disk. 3. Each primordial ring is associated with the local mid-plane temperature $T$ of the disk. 4. Each of these $T$'s is related to a photon energy ($E_p$) in the portion of the $H_2$ spectrum in Table 4. 5. The relationship between $T$ and $E_p$ is linear and of the form

$$T = C_1(E_p + C_2), \qquad (3)$$

where $C_1$ and $C_2$ are constants empirically determined in section 2.3.c.

Section 2.6 and Appendix 3 theoretically link $E_p$'s to specific disk temperatures ($T$'s) in the linear form of Eq. (3).

### 2.3.c Determination of $C_1$ and $C_2$

Mousis (2004) derives a series of TD's for a "water-rich" Uranian protosatellite disk. Each TD is for a different time $t$ after the disk is formed by an assumed impact to Uranus's surface. Fig. 1 of Mousis (2004) has seven TD's with $t$ values for the first three equal to $1 \times 10^4$, $2 \times 10^4$ and $5 \times 10^4$ years. Fig. 6 in the present investigation includes reproductions of the first three Mousis TD's (each represented by solid straight line segments) and two TD's (dashed line segments) constructed in the present investigation by using the similarities among the first three Mousis TD's. Appendix 1 is a discussion of the method used for the reproduction of the Mousis TD's and for the construction of the two dashed TD's. The $t$ values for TD's are indicated in Fig. 6. Three of these are from Mousis (2004) and the other two associated with the dashed TD's are $t$ = 8800 and 6900 years. A discussion of how these two $t$ 's are determined is in Appendix 2.

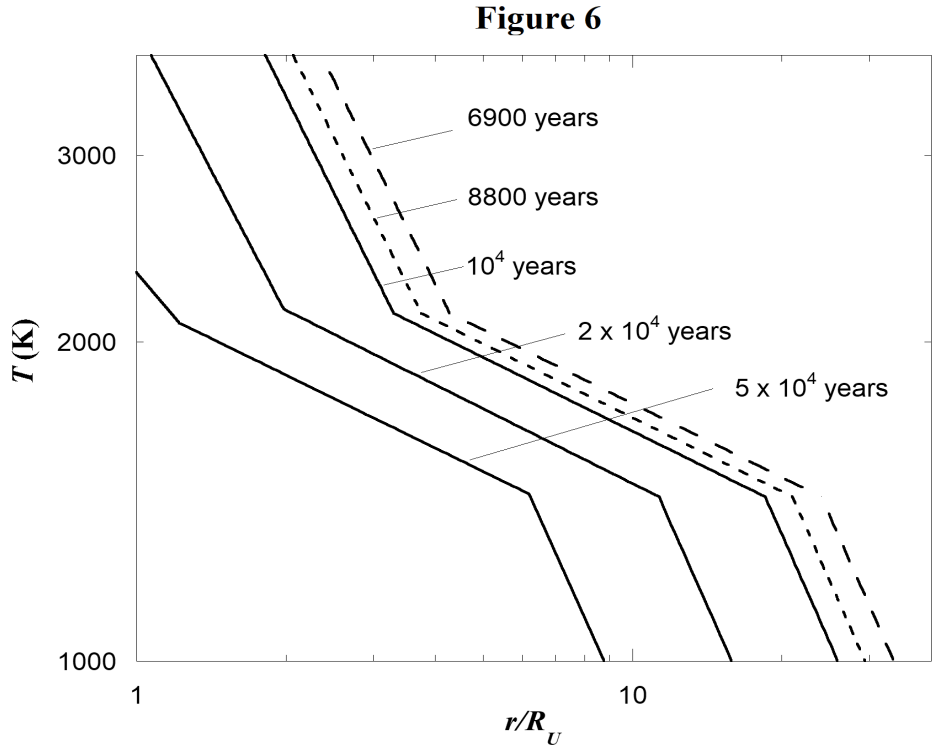

**Fig. 6. TD's with solid lines: Mousis (2004)
TD's with dashed lines: present investigation**

**Figure 6**

Table 4 and the time dependent series of TD's in Fig. 6 are used to determine the parameters $C_1$ and $C_2$ in Eq. (3). Because graphs in Fig. 6 do not contain a dip and a peak as is seen in the Uranian PED in Fig. 5a, the only $E_p$'s and $R_{ui}$'s used in the process are those in the PED's "tail" for which $i = 9 - 20$ in the second page of Table 4. This parameters are found as follows. Many trial sets of $C_1$ and $C_2$ are used in Eq. (3). For each set of $C_1$ and $C_2$, a set of $T$'s is calculated with each $T$ in a set corresponding to an $E_p$ and a $R_{ui}$ in Table 4. Then each set of $T$'s and $R_{ui}$'s is graphed to make a trial TD and the trail TD is compared to the TD's in Fig. 6.

Best fits to the Mousis (2004) TD's for which $t = 1 \times 10^4$, $2 \times 10^4$ and $5 \times 10^4$ years are found and the best of the three is the one for the earliest time $t = 1 \times 10^4$ years shown in Fig. 7. But it is seen that a Mousis type TD for an earlier time would most likely be fitted even better. Various earlier time Mousis type TD's are constructed using a method discussed in Appendix 1. By trial and error it is found that the TD corresponding to $t$ = 8800 years is fitted best of all. This TD and the fitted data are given in Fig. 8. The corresponding parameters are $C_1$ = 2.315 K·cm and $C_2$ = -3720 cm$^{-1}$. In the present model, the time $t = t_0$ = 8800 years is presumably the approximate time when the Uranian satellites start their formation. The best fit to the TD for $t$ = 6900 years is given in Fig. 9. It is interesting to compare Figs. 7, 8 and 9. The times $t_0$ = 8800 and $t$ = 6900 years are determined in Appendix 2.

**Fig. 7. Line segments: The TD for $t = 1 \times 10^4$ years (Mousis (2004).
The point wise TD is determined by varying $C_1$ and $C_2$ in Eqn. (3).**

Figure 7

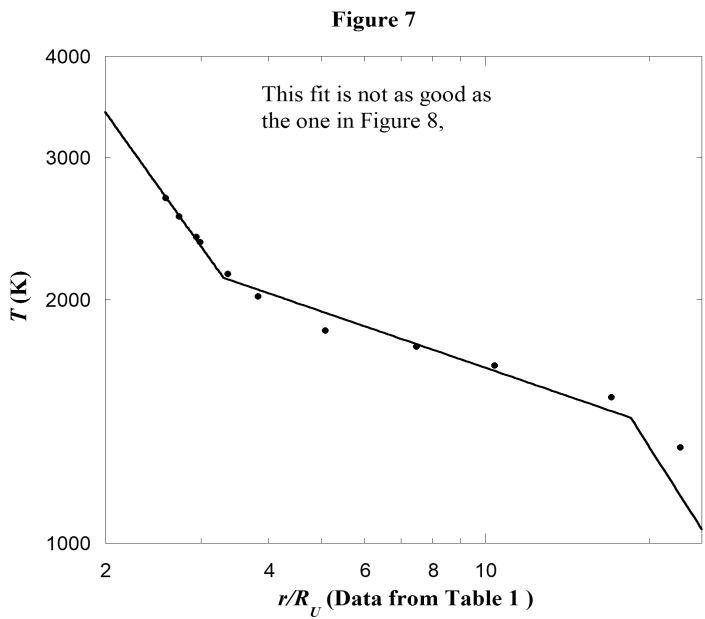

**Fig. 8.** Line segments: The Mousis "type" TD for $t = t_0 = 8800$ years. The point wise TD is determined by varying $C_1$ and $C_2$ in Eqn. (3).

**Figure 8**

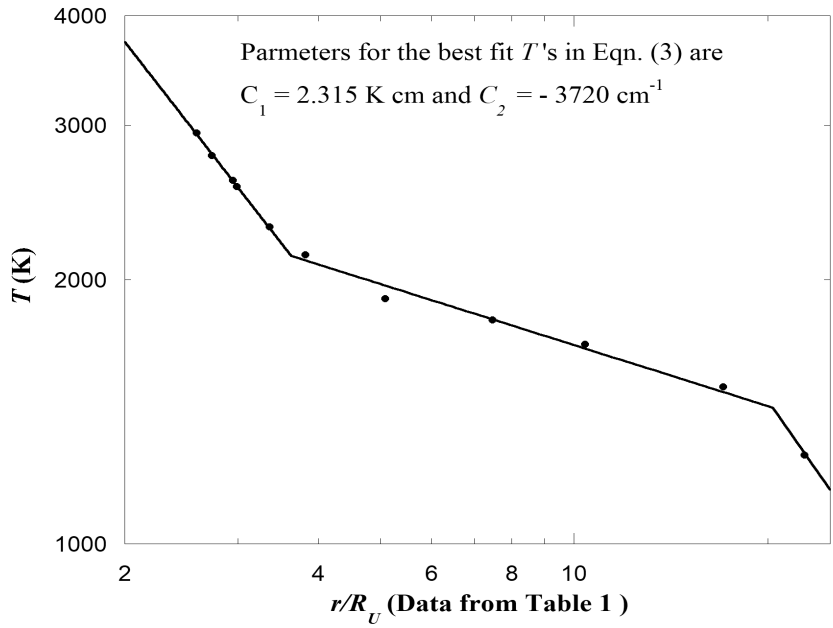

Parameters for the best fit $T$'s in Eqn. (3) are
$C_1 = 2.315$ K cm and $C_2 = -3720$ cm$^{-1}$

**Fig. 9.** Line segments: The Mousis "type" TD for $t = t_0 = 6900$ years. The point wise TD is determined by varying $C_1$ and $C_2$ in Eqn. (3).

**Figure 9**

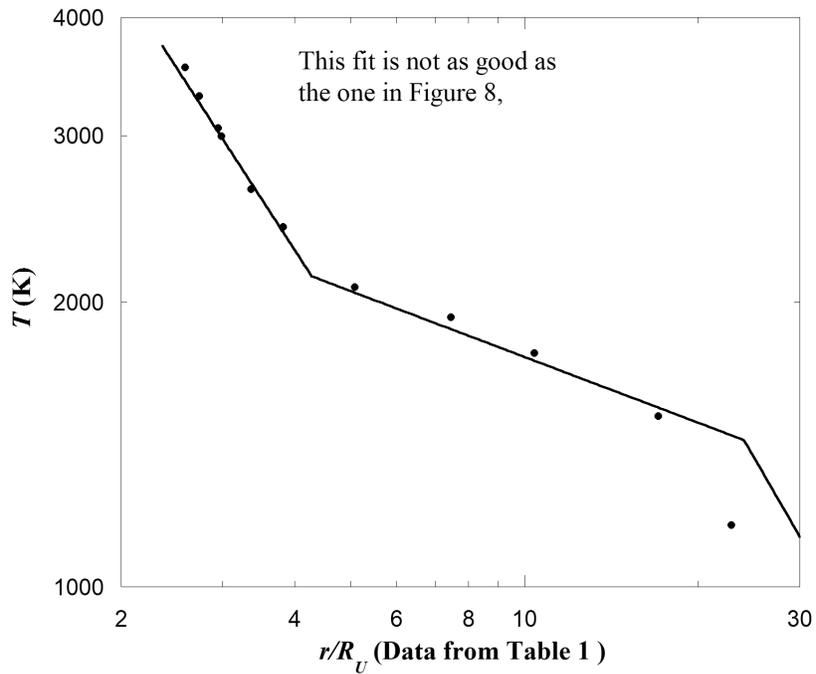

This fit is not as good as the one in Figure 8,

## 2.4. The Overlap of the Uranian, Jovian and Neptunian TD's

Table 2 contains transformed orbital radii of the Jovian and Neptunian satellites. It is used in subsection 2.1.c to construct Fig. 3. Now we use it again to construct the scaled Jovian and Neptunian TD's seen in Fig. 10. This figure also includes the graphs shown in Fig. 8 for the Uranian TD for which $t = t_0 = 8800$ years. Fig. 10 shows the generally precise overlap of the three TD's. As discussed in section 2.1.c, this overlap cannot continue inward to smaller orbital radii because the Jovian and Neptunian protosatellite disks are relatively closer to their respective planets than is the Uranian disk from Uranus. Interestingly, in the upper left the shape of the Uranian and Neptunian TD's both show a "shoulder". In a future discussion we see a similar shoulder in Saturn's TD.

**Fig. 10. Radial scaling causes the Uranian, Jovian and Neptunian TD's to overlap.**

**Figure 10**

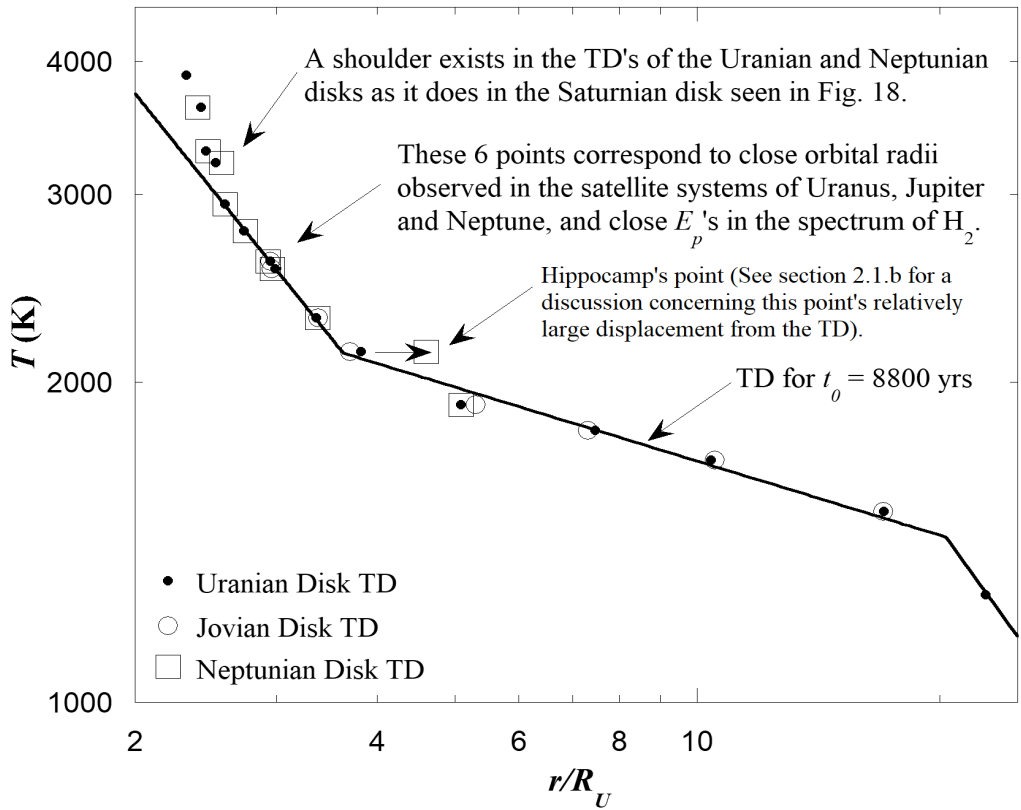

A shoulder exists in the TD's of the Uranian and Neptunian disks as it does in the Saturnian disk seen in Fig. 18.

These 6 points correspond to close orbital radii observed in the satellite systems of Uranus, Jupiter and Neptune, and close $E_p$'s in the spectrum of $H_2$.

Hippocamp's point (See section 2.1.b for a discussion concerning this point's relatively large displacement from the TD).

TD for $t_0 = 8800$ yrs

- Uranian Disk TD
- Jovian Disk TD
- Neptunian Disk TD

In the present model Uranian satellites begin their formation approximately 8800 years after the the Uranian protosatellite disk was formed. Because the Jovian and Neptunian TD's overlap the Uranian TD for which $t = 8800$ years, we may think it is possible to draw conclusions about the times when the Jovian and Uranian satellites begin their formation. But this is not true. For the most part, TD's corresponding to different times in the vicinity of $t = 10^4$ years can be transformed from one to another simply by linearly scaling the radial coordinate system associated with one of them. So Eqs. (1) and (2) scale TD's with respect to both time and space but we don't know to what degree for each.

Table 5. Photon energies ($E_p$'s) are used to calculate the Uranian protosatellite disk temperatures ($T$'s). The $T$'s and orbital radii ($R_u$'s) are used to construct the complete Uranian TD in Fig.11. All T's are calculated from Eq. (3) with $C_1$ = 2.315K·cm and $C_2$ = -3720 cm$^{-1}$.

| Satellite | $i$ | $E_p$(cm$^{-1}$)[a] | T(K) | $R_u$[b] |
|---|---|---|---|---|
| Ring 6 | 11&12 | 4826 | 2560 | 1.637 |
| Ring 5 | 13&14 | 4712 | 2296 | 1.652 |
| Ring 4 | 15 | 4642 | 2134 | 1.666 |
| Ring α | 15 | 4642 | 2134 | 1.750 |
| Ring β | 13&14 | 4712 | 2296 | 1.786 |
| Ring η | 11&12 | 4826 | 2560 | 1.846 |
| Ring γ | 10 | 4917 | 2771 | 1.863 |
| Ring δ | 9 | 4990 | 2939 | 1.900 |
| Cordelia | 7 | 5108 | 3213 | 1.948 |
| Ring λ | 5&6 | 5144 | 3296 | 1.957 |
| Ring ε | 3&4 | 5285 | 3623 | 2.006 |
| Ophelia | 1 | 5397 | 3882 | 2.105 |
| Bianca | 1 | 5397 | 3882 | 2.316 |
| Cressida | 3&4 | 5285 | 3623 | 2.418 |
| Desdemona | 5&6 | 5144 | 3296 | 2.453 |
| Juliet | 7 | 5108 | 3213 | 2.520 |
| Portia | 9 | 4990 | 2939 | 2.586 |
| Rosalind | 10 | 4917 | 2771 | 2.735 |
| Cupid | | | | 2.911 |
| Belinda | 11 | 4841 | 2596 | 2.946 |
| Perdita | 12 | 4823 | 2553 | 2.990 |
| Puck | 13&14 | 4712 | 2296 | 3.365 |
| Mab | 15 | 4642 | 2134 | 3.824 |
| Miranda | 16 | 4543 | 1904 | 5.082 |
| Ariel | 17 | 4498 | 1800 | 7.469 |
| Umbriel | 18 | 4449 | 1688 | 10.410 |
| Titania | 19 | 4372 | 1510 | 17.070 |
| Oberon | 20 | 4265 | 1262 | 22.830 |

[a]Black van Dishoeck (1987). Some $E_p$'s are weighted averages.
[b]NASA (2021)

## 2.5. The Complete TD for the Uranian Protosatellite Disk

Eq. (3) relates $T$'s to $E_p$'s. The constants in this relationship are determined in section 2.3.c by fitting data to the "tail" of the Uranian disk TD. We now assume Eq. (3) holds for the complete Uranian disk TD. Table 5 lists the complete set of calculated $T$'s, and Fig. 11 shows the complete point-wise TD. Because the relationship between $T$ and $E_p$ in Eq. (3) is linear, the Uranian TD and PED have similar shapes. Therefore the Uranian disk TD also shows characteristics described by L&P (1985) for the solar nebula as described in section 2.3.a. Notice the result from the present investigation diverges from the Mousis "type" graph at a point just to the right of the peak. A similar result is seen later in Fig. 18 where two Saturn TD's are compared.

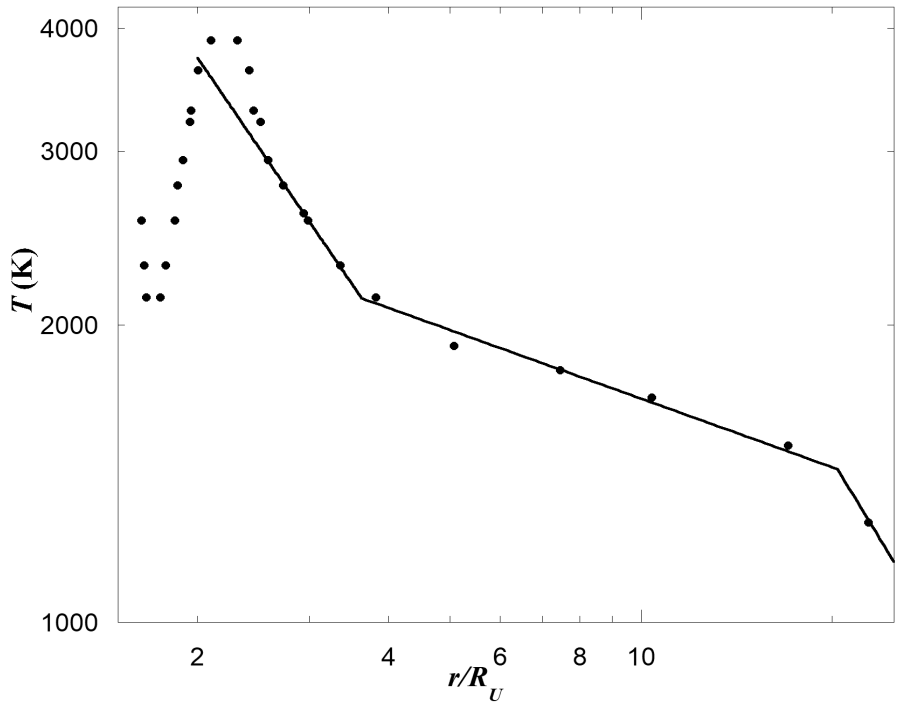

Fig. 11. The complete Uranian disk TD overlapping the Mousis (2004) "type" TD ($t = t_0 = 8800$ years)

Figure 11

## 2.6. Why are T and $E_p$ related in the Form Given by Eq. (3)?

Stancil and Dalgarno (1997) have used the mechanism of SRMA to study rate coefficients for the production of the molecule LiH from the atoms Li and H in the early universe. The form of the reaction describing SRMA is

$$A + B + h\nu \rightarrow AB + 2h\nu, \qquad (4)$$

where $A$ and $B$ are atoms or molecules that associate to make the molecule $AB$. The photon $h\nu$ on the left side of Eq. (4) stimulates the process during which a second photon is created with the same energy and direction of motion as the first. And so there are two photons symbolized on the right side.

In this section, the SRMA mechanism is used to explain why orbital radii of satellites are related to photon energies in the $H_2$ spectrum. Consider a protosatellite disk consisting of many colliding atoms or molecules. We focus on collisions that occur near the mid-plane of the disk. The types of atoms or molecules are not determined (nor do they need to be) in this analysis and we simply call them *A, B* and *AB*.

Resonance is established at just certain radii, corresponding to what we call resonance rings, in the protosatellite disk wherever there is a "match" between local thermal energy in the disk and photon energies belonging to photons in a discrete spectrum such as the spectrum of molecular hydrogen. The following model is true for SRMA reactions that contribute to resonance. For this model assume these reactions are viewed from a coordinate system that is in corotation with the gas. Also, the possibility that the source of photons is the protosun is mentioned in subsection 2.2.a. If true, then resonance is most likely initiated during the two time periods during the protoplanet's orbit around the protosun when the protoplanet's equatorial plan is essentially parallel to the direction of the protosun's radiation.

1. SRMA reactions happen in abundance due to the chain reaction that results from the multiplying effect of the additional photon produced with each reaction. Because photons are created at the expense of kinetic energy, the temperature decreases in resonance rings.

2. On average, during a resonant reaction the kinetic energy of *A* equals the kinetic energy of *B* for all resonant reactions. This average energy is proportional to *T* and is called $K_{mp}$ the **most probable kinetic energy** for *A* and *B* in the distribution of energies in the gas. In the present model $K_{mp}$ is taken to be *kT*, where *k* is the Boltzmann constant. More is said about $K_{mp}$ in Appendix 3.

3. During an SRMA reaction a photon (*hv*) is formed (See Eq. (4)). There are two contributions to the photon's energy. i.e. $E_p = |\Delta K| + \Delta$, where $\Delta K$ is the loss of kinetic energy of *A* and *B* due to their collision and association to form the molecule *AB* and $\Delta$ is a transition energy that *AB* may undergo as it is being created.

4. In a resonance ring within a protosatellite disk, after the collision the photons (*hv's*) and molecules (*AB's*) move essentially tangent to the ring. This condition ensures that photons continue to cause resonant reactions for long distances until they are no longer within the ring. It also ensures *AB*'s are collecting within a narrow ring at a temperature lower than *T*. The direction of the formed molecules is the same as the direction of the circulating gas. The direction of the resonant photons is parallel or antiparallel to the direction of the circulating gas.

5. Once resonance is initiated it persists even though the temperature in the resonance ring becomes increasing lower. At any gas temperature there are always *A*'s and *B*'s with kinetic energies equal to *kT* where *T* is the temperature when the resonance starts.

In subsection 2.3.c the Mousis (2004) *TD*'s were used to determine the constant $C_1$ = 2.315 K·cm in Eq(3) which is rewritten as Eq. (6) below. Using the above model a theoretical relationship between *T* and $E_p$ for resonance in the disks of Uranus, Jupiter and Neptune is determined in Appendix 3 with the result:

$$T = (hc/k)(E_p - \Delta) \qquad (5)$$

where $hc/k$ = 1.4388 K·cm, $\Delta$ is the energy defined in the model above, *h* is Planck's constant and c is the speed of light. Eq. (5) is compared to the empirically determined relationship

$$T = (2.315 \text{ K·cm})(E_p - 3720 \text{ cm}^{-1}), \qquad (6)$$

or
$$T = 1.609(hc/k)(E_p - 3720 \text{ cm}^{-1}), \tag{6a}$$

where $1.609(hc/k) = C_1 = 2.315$ K·cm.

We note the reasonable agreement between the multiplying constants (1 and 1.609) on the right sides of Eqs. (5) and (6a). This agreement supports the usefulness of the present model and the technique used to determine $C_1$. Possibly if the simplifying approximations made in Appendix 3 were relaxed, the agreement would be better. The constant 3720 cm$^{-1}$ in Eq. (6a), corresponding to 0.461 eV, may be helpful in the identification of $A$ and $B$.

### 2.7. The Photon Energy Distribution in Saturn's Protosatellite Disk

The complicated Saturnian ring and satellite system cannot be fit using the $H_2$ spectrum, and so we turn to the hydrogen atom ($H$) spectrum as an alternative. Again, the origin of the $H$ photons is not identified in the present investigation. But as before, perhaps the source of the radiation is the protosun as a T Tauri star (Kwan and Fischer 2011) possibly in a FUor or Exor state (Hartmann, Kenyon and Hartigan 1993). A model similar to the one put forward in this paper is used to investigate the origin of the planets and other satellites of the Sun (Lombardi 2015b). Results of that investigation suggest that the source of the required $H$ radiation is the protosun in a FUor or Exor state.

The Rydberg formula gives each transition energy $E_p(n_f, n_i)$ for the hydrogen atom in terms of the principle quantum numbers, $n_i$ and $n_f$. In the Bohr model these quantum numbers correspond to the initial and final states of the electron for a particular transition. The $H$ spectrum contains many series of photon energies (each series corresponds to a different $n_f$ value) with each series possessing a distinct limit $E_p(n_f, \infty)$. As each $n_i$ approaches $\infty$ an edge in the $H$ spectrum is also approached. Also, Saturn's broad rings have edges, some of which correspond to edges in the $H$ spectrum.

The spacing between energies in the $H$ spectrum approaches zero as a series limit is approached. And because resonance rings have finite widths, they overlap to a large degree near some edges of Saturn's rings. On the other hand some of Saturn's rings have edges that don't correspond to a series limit. As $n_i$ gets smaller the spacing between energies gets larger. Resonance rings overlap less and less until they no longer overlap producing another type of ring edge. Outside this type of ring edge, resonance rings create narrow primordial rings that evolve into satellites.

Assuming the hydrogen spectrum of $E_p$'s creates the key resonances in Saturn's protoplanetary disk, we can determine which $E_p$ is associated with each of the satellites of Saturn by correctly pairing some series limits with certain ring edges. Lombardi (2015a) reports a first attempt at making this association of series limits with ring edges that was somewhat successful from the inner edge of the A ring out to Titan. But it was not able to pair hydrogen $E_p$'s with any satellites or rings orbiting closer to Saturn's surface. The present investigation improves the association of limits and edges and successfully explains the complicated ring and satellite system of Saturn.

Table 6 gives the energies of photons emitted during transitions in the range 617.3-2239.5 cm$^{-1}$, where energy is in wave numbers (cm$^{-1}$). In the present model, this range of $E_p$'s is associated with the A and E rings and all the satellites in their vicinity. When the inner edges of the A and E rings of Saturn are associated with the series limits $E_p(7,\infty)$ and $E_p(8,\infty)$ the photons in this range are automatically paired with satellite orbital radii or connected to the A or E rings. These findings lead to the development of a complete PED that is similar in shape to the PED shown in Fig. 5a for all of Uranus's protosatellite disk. (I.e. both have a dip, a peak and a "tail"). In section 2.8, a TD for Saturn's protosatellite disk is determined that matches a theoretically determined Saturnian disk TD (Mousis, Gautier and Bockelee-Moran 2002) in the "tail" of the TD. In Tables 6 and 7 and elsewhere, integers with square brackets are indices that identify satellites and some ring edges.

Table 6. Photon energies ($E_p(n_f,n_i)$) in the hydrogen spectrum used to construct Table 7 and Fig. 12. $E_p$'s are in units of cm$^{-1}$. Bolded $E_p$'s are in the range 617.3-2239.5 cm$^{-1}$. $n_f$ and $n_i$ are defined in section 2.7.

| $n_i$ | $n_f=4$ | [$i$] | $n_f=5$ | [$i$] | $n_f=6$ | [$i$] | $n_f=7$ | [$i$] | $n_f=8$ | $n_i$ |
|---|---|---|---|---|---|---|---|---|---|---|
| 5 | (High $E_p$'s) | | x | | x | | x | | x | 5 |
| 6 | ↓ | [15] | *1341.2* | | x | | x | | x | 6 |
| 7 | | [2] | *2149.9* | [18] | *808.7* | | x | | x | 7 |
| 8 | | | (High $E_p$'s) | [15] | *1333.6* | | 524.9 | | x | 8 |
| 9 | | | ↓ | [11] | *1693.5* | [17] | *884.8* | | 359.9 | 9 |
| 10 | | | | [4] | *1950.9* | [16] | *1142.2* | [19] | *617.3* | 10 |
| 11 | | | | [2] | *2141.3* | [15] | *1332.6* | [18] | *807.7* | 11 |
| 12 | | | | | (High $E_p$'s) | [14] | *1477.5* | E ring | 952.6 | 12 |
| 13 | | | | | ↓ | [13] | *1590.2* | E ring | 1065.3 | 13 |
| 14 | | | | | | [12] | *1679.6* | E ring | 1154.8 | 14 |
| 15 | | | | | | [9] | *1751.8* | E ring | 1226.9 | 15 |
| 16 | | | | | | [8] | *1810.9* | E ring | 1286.0 | 16 |
| 17 | | | | | | [7] | *1859.8* | E ring | 1334.9 | 17 |
| 18 | | | | | | [6] | *1900.8* | E ring | 1375.9 | 18 |
| 19 | | | | | | [5] | *1935.6* | E ring | 1410.7 | 19 |
| 20 | | | | | A ring&[3][1] | | *1965.2* | E ring | 1440.3 | 20 |
| 21 | | | | | A ring | | 1990.7 | E ring | 1465.8 | 21 |
| 22 | | | | | A ring | | 2012.8 | E ring | 1487.9 | 22 |
| 23 | | | | | A ring | | 2032.1 | E ring | 1507.2 | 23 |
| 24 | | | | | A ring | | 2049.0 | E ring | 1524.1 | 24 |
| 25 | | | | | A ring | | 2064.0 | E ring | 1539.1 | 25 |
| ↓ | | | | | | | ↓ | | ↓ | ↓ |
| ∞ | | | | | IE A[1] | | *2239.5* | IE E[10] | *1714.6* | ∞ |

IE stands for Inner Edge.
Bolded $E_p$'s are assigned to individual satellites and some ring edges in Saturn's satellite system.
Generally unbolded $E_p$'s contributed to the creation of the A ring and E ring. IE A is inner edge A ring.
A satellite index [$i$] to the left of each bolded $E_p$ is assigned to a satellite or ring edge in Table 7.
All unbolded close $E_p$'s contribute to either the A or E ring of Saturn.
[1] Daphnis ($E_p(7,20)$,[3]) is in the Keeler Gap near the outer edge of the A ring. NASA(2021)
$E_p(7,8)$ and $E_p(8,9)$ are out of the range of interest.
(High $E_p$'s) is indicated for many $E_p$'s because they are out of range of interest.
$E_p$'s corresponding to $n_f=9$ or larger are not included. Apparently they did not create resonance.

### 2.7.a. The PED for Saturn's A Ring and Beyond

Table 7 lists the inner and outer edges of the A, G and E rings and satellites from Pan to Titan. Hyperion lies just beyond Titan with an orbital inclination and orbital eccentricity that classify it as regular. However it has an irregular shape and it is believed that Hyperion is a remnant of a larger satellite that experienced a catastrophic impact (Farinella, Marzari, and Matteoli 1997). Iapetus is beyond Hyperion and the inclination of its orbital plane is 14.7°. All the other regular satellites of Saturn have inclinations of their planes near 0°. Perhaps Iapetus also experienced an impact or it was captured. Hyperion and Iapetus are not included in the present analysis.

Table 7. Photon energies ($E_P$'s) and orbital radii ($r/R_s$) used for the construction of Fig. 12. The individual pairings of $E_p$'s and $r/R_s$'s are determined by first pairing $E_p(7,\infty)$ and $E_p(8,\infty)$ with the inner radii of the A and E rings respectively. Then the other pairings automatically fall into place.

| Satellite or Ring Edge Name | $[i]$[a] | $r/R_s$[b] | $n_f, n_i$[c] | $E_p(n_f, n_i)$[c] (cm$^{-1}$) |
|---|---|---|---|---|
| Inner Edge A ring | [1] | 2.030 | 7,∞ | 2239.5 |
| Pan in Encke Gap | [2] |  | 5,7 | 2149.9 |
| Pan in Encke Gap | [2] |  | 6,11 | 2141.4 |
| Average of Pan $E_p$'s | [2] | 2.217 |  | 2145.7 |
| Daphnis in Keeler Gap | [3] | 2.265 | 7,20 | 1965.2 |
| Outer Edge A ring |  | 2.270 |  |  |
| Atlas | [4] | 2.284 | 6,10 | 1950.9 |
| Prometheus | [5] | 2.312 | 7,19 | 1935.6 |
| F ring | [6] | 2.320 | 7,18 | 1900.8 |
| Pandora | [7] | 2.351 | 7,17 | 1859.8 |
| Epimetheus and Janus | [8] | 2.511 | 7,16 | 1810.9 |
| Inner Edge G ring |  | 2.754 |  |  |
| Aegaeon | [9] | 2.779 | 7,15 | 1751.8 |
| Outer Edge G ring |  | 2.871 |  |  |
| Inner Edge E ring | [10] | 2.987 | 8,∞ | 1714.6 |
| Mimas | [11] | 3.073 | 6,9 | 1693.5 |
| Methone | [12] | 3.219 | 7,14 | 1679.6 |
| Anthe | [13] | 3.280 | 7,13 | 1590.2 |
| Pallene | [14] | 3.501 | 7,12 | 1477.5 |
| Enceladus | [15] |  | 5,6 | 1341.2 |
| Enceladus | [15] |  | 6,8 | 1333.6 |
| Enceladus | [15] |  | 7,11 | 1332.6 |
| Average of Enceladus $E_p$'s | [15] | 3.949 |  | 1335.8 |
| Tethys, Calypso & Telesto | [16] | 4.889 | 7,10 | 1142.2 |
| Dione, Helene & Polydeuces | [17] | 6.262 | 7,9 | 884.8 |
| Outer Edge E ring |  | 7.964 |  |  |
| Rhea | [18] |  | 6,7 | 808.7 |
| Rhea | [18] |  | 8,11 | 807.7 |
| Average of Rhea $E_p$'s | [18] | 8.745 |  | 808.2 |
| Titan | [19] | 20.273 | 8,10 | 617.3 |

[a] The satellite indices ($[i]$ values) are assigned to satellites in this table and used in Table 6 and Fig. 12.

[b] Orbital radii in units of the equatorial radius of Saturn. NASA(2021) As discussed in section 2.8, the orbital radii ($r/R_s$) are transformed with Eq. (8) to give these transformed orbital radii ($r/R_s$)' used in Figs. (17) and (18).

[c] The quantum numbers that define transitions in the hydrogen atom and photon energies associated with these transition.

In Table 6 the $n_f = 7$ and $n_f = 8$ series contain close unbolded $E_p$'s where $n_i$ is large at or near the series limits $E_p(7,\infty)$ and $E_p(8,\infty)$. In these regions resonance rings overlap each other and create a broad circular ring where resonance occurs thus laying the ground work for the A and E rings of Saturn. The outer edge of the E ring corresponds to where resonance rings no longer overlap between $E_p(8,11)$ and $E_p(8,12)$. The energy $E_p(8,11)$ corresponds to the first satellite outside the E ring where the resonance ring is not wide enough to overlap an adjacent resonance ring. This satellite is Rhea ($[i] = 18$). The bolded $E_p$'s at the very bottom of these columns are the series limits $E_p(7,\infty) = 2239.5$ and $E_p(8,\infty) = 1714.6$ cm$^{-1}$. These limits and the indices, [1] and [10], are associated with the inner edges of the A ring and E ring respectively. The G ring's inner edge is not associated with a series limit. Its existence is likely related to an arc of debris that exists near the orbit of the satellite Aegaeon (Hedman et al. 2007a).

In Table 7 the $r/R_s$ values are the orbital radii of satellites and ring edges in units of $R_s$ the equatorial radius (60,268 km) of Saturn. Except for the outer edges of the A and E rings and the inner and outer edges of the G ring, these features are each assigned a satellite index $[i]$. All together there are 23 bolded $E_p$'s in Table 6. The close $E_p$'s for $[i] = 2$ (Pan) are averaged as shown in Table 7. The same procedure holds for $[i] = 15$ and 18. With close $E_p$'s accounted for, the satellite indices range from [1] to [19]. This number of indices accommodates the 19 satellites and ring inner edges (not including the G rings inner edge) from the inner edge of the A ring to Titan. In Tables 6 and 7, notice the outer edge of the A ring occurs very close to the orbital radius of Daphnis with $E_p = E_p(7,20)$. There are no $E_p$ values listed in Table 7 between $E_p(7,20)$ and $E_p(7,\infty)$ because they all are within the A ring. On the other hand outside the A ring $E_p$ values from $E_p(7,20)$ to $E_p(7,9)$ are all assigned to satellites or the narrow F ring. A similar result can be seen for the E ring. Interestingly all the remaining satellites in the region are accommodated by other $E_p$'s belonging to other series with out any leftover $E_p$'s or leftover satellites. This is a major success of the recent model.

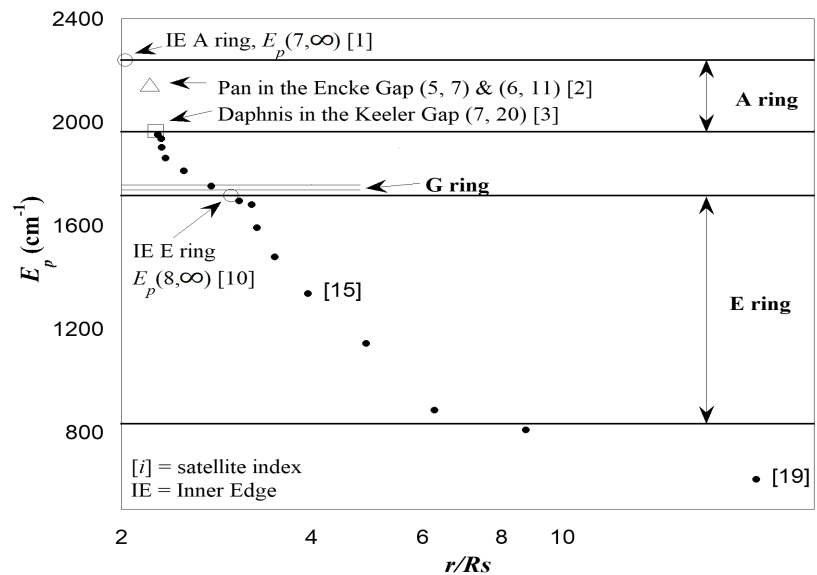

Fig. 12. The Photon Energy Distribution in Saturn's Protosatellite disk for the A ring and beyond

Figure 12

In Table 6 consider $E_p(8,10) = 617.3$ cm$^{-1}$. This is the lowest $E_p$ value assigned to a satellite (Titan with $i = [19]$). At the top of the columns labeled $n_f = 7$ and $n_f = 8$, there are two unbolded $E_p$'s, $E_p(7,8) = 524.9$ cm$^{-1}$ and $E_p(8,9) = 359.9$ cm$^{-1}$ that are less than $E_p(8,10)$. If there were two more regular satellites beyond Titan to be considered, they would be associated with these $E_p$'s. There are many $E_p$'s in Table 6 labeled High $E_p$. They are all larger than $E_p(7,\infty) = 2239.5$ cm$^{-1}$, the upper range of $E_p$'s in Table 7. $E_p$'s corresponding to $n_f \geq 9$ are not included in the analysis. Apparently they did not create resonance.

Fig. 12 is constructed using the data in Table 7. This figure is the PED for Saturn's protosatellite disk from the inner edge of the A ring out to Titan.

*2.7.b. The PED for Saturn's D, C, B and A Rings*

Now consider the PED for Saturn's protoplanetary disk from the inner edge of the D ring out to the outer edge of the A ring. To proceed we explore the possibility that Saturn's disk PED has the same general shape close to Saturn's surface as does Uranus's disk PED close to Uranus's surface. I.e. its shape contains a dip and a peak similar to what are seen in Figs. 5a and 5b. If it does, then some particular $E_p$'s could contribute to Saturn's disk PED up to three times. For this reason it is useful to study the photograph "Panoramic scan across Saturn's rings" (Planetary Society 2021). This scan has labels on distinct gaps in Saturn's rings. Notice the similar appearance of the Colombo, Maxwell and Encke Gaps. These three gaps are among the six widest gaps within Saturn's rings. (The other three are in the Cassini Division (Planetary Society 2021)). The Titan ringlet[24], Maxwell ringlet[23] and Pan[2] are between the boundaries of each gap respectively (NASA 2021). These similarities lead to the assignment of the same $E_p$'s to the two ringlets as are already assigned to Pan with index [2] in Table 7. To better see this, consider the set of indices [2,23,24] in two places in Tables 8 and the indices [2], [23] and [24] in Table 9. Each of the two ringlets and Pan are associated with *two* closely spaced $E_p$'s, $E_p(5,7)$ and $E_p(6,11)$. Perhaps the two support each other creating an extra strong resonance within the boundary of each gap thus creating the gaps and causing the material within the gaps to collect in resonance rings that evolve into the two ringlets and Pan. Similar considerations lead to the assignment of $E_p(7,\infty)$ to the inner and outer edges of the C ring ([$i$] =[22] and [25]) just as it is assigned to the inner edge of the A ring ([$i$] =[1]). The shape of the PED's dip in Fig.13 is therefore defined by the four points labeled [22], [23], [24] and [25].

In Table 8 the series of close $E_p$'s from $E_p(7,20)$ to $E_p(7,\infty)$ and from $E_p(7,29)$ to $E_p(7,\infty)$ create resonances that produce the A ring and C ring respectively. Also the series of close $E_p$'s from $E_p(6,12)$ to $E_p(6,\infty)$ produces the D ring. The C ring is different from the A, D and E rings in that the C ring is centered at the center of the dip in the PED in the Saturnian disk. Therefore *both* of the C ring's edges are associated with a photon series limit, $E_p(7,\infty)$. Also, because the B ring is at the peak in the PED, neither of its edges are associated with a photon series limit.

The PED in Fig.13 rises upward from its minimum towards Saturn's surface where it ends at the inner edge of the D ring [29] where $E_p = E_p(6,\infty)$. The radii of the ringlets D68, D72 and D73 in Saturn's D ring are determined (Hedman et al. 2007b) from Cassini spacecraft images. These ringlets, with indices [26], [27] and [28], and the inner radius of the D ring [29] are the distinctive features of the ring.

Table 8. Bolded $E_p$'s are in the range 1965.2-3048 cm$^{-1}$. These $E_p$'s are used to construct Table 9 and Figs.13 and 14.

| $n_i$ | Comments | $n_f = 4$ | Comments | $n_f = 5$ | Comments | $n_f = 6$ | Comments | $n_f = 7$ |
|---|---|---|---|---|---|---|---|---|
| 5 | Inner&OuterBring&[26] | **2469.1** | | | | | | |
| 6 | (high $E_p$'s) | 3810.3 | (low $E_p$) | 1341.2 | | | | |
| 7 | ↓ | 4619.0 | [2, 23, 24] | **2149.9** | (low $E_p$) | 808.7 | | |
| 8 | | 5143.9 | [27] | **2674.8** | (low $E_p$) | 1333.6 | (low $E_p$'s) | 524.9 |
| 9 | | 5503.8 | [28] | **3034.7** | (low $E_p$) | 1693.5 | ↓ | 884.8 |
| 10 | | 5761.2 | (high $E_p$'s) | 3292.1 | (low $E_p$) | 1950.9 | | 1142.2 |
| 11 | | 5951.6 | ↓ | 3482.6 | [2, 23, 24] | **2141.3** | | 1332.6 |
| 12 | | 6096.5 | | 3627.4 | Dring&InnerBring&[20] | **2286.2** | | 1477.5 |
| 13 | | 6209.2 | | 3740.1 | Dring&InnerBring&[21] | **2398.9** | | 1590.2 |
| 14 | | 6298.7 | | 3829.6 | Dring&Inner&OuterBrings | 2488.4 | | 1679.6 |
| 15 | | 6370.8 | | 3901.8 | Dring&Inner&OuterBrings | 2560.5 | | 1751.8 |
| 16 | | 6429.9 | | 3960.8 | D ring | 2619.6 | | 1810.9 |
| 17 | | 6478.8 | | 4009.8 | D ring | 2668.5 | | 1859.8 |
| 18 | | 6519.9 | | 4050.8 | D ring | 2709.6 | ↑ | 1900.8 |
| 19 | | 6554.6 | | 4085.5 | D ring | 2744.3 | (low $E_p$'s) | 1935.6 |
| 20 | | 6584.2 | | 4115.1 | D ring | 2773.9 | A ring [3][1] | **1965.2** |
| 21 | | 6609.7 | | 4140.6 | D ring | 2799.4 | A ring | 1990.7 |
| 22 | | 6631.8 | | 4162.8 | D ring | 2821.5 | A ring | 2012.8 |
| 23 | | 6651.1 | | 4182.0 | D ring | 2840.8 | A ring | 2032.1 |
| 24 | | 6668.0 | | 4199.0 | D ring | 2857.7 | A ring | 2049.0 |
| 25 | | 6683.0 | | 4213.9 | D ring | 2872.7 | A ring | 2064.0 |
| 26 | | 6696.2 | | 4227.1 | D ring | 2885.9 | A ring | 2077.2 |
| 27 | | 6708.0 | | 4238.9 | D ring | 2897.7 | A ring | 2089.0 |
| 28 | | 6718.6 | | 4249.5 | D ring | 2908.3 | A ring | 2099.6 |
| 29 | | 6728.1 | | 4259.0 | D ring | 2917.8 | A&C rings | 2109.0 |
| 30 | | 6736.6 | | 4267.6 | D ring | 2926.3 | A&C rings | 2117.6 |
| 31 | ↑ | 6744.4 | ↑ | 4275.3 | D ring | 2934.1 | A&C rings | 2125.3 |
| 32 | (high $E_p$'s) | 6751.4 | (high $E_p$'s) | 4282.3 | D ring | 2941.1 | A&C rings | 2132.4 |
| ∞ | ( high $E_p$) | 6858.6 | high $E_p$ | 4389.5 | [29] | **3048.3** | [ 1, 22, 25] | **2239.5** |

[1]Daphnis is in the Keeler Gap near the outer edge of the A ring. NASA(2021)

Notes: $E_p$'s used for Fig. 13. are in the range (1965.2-3048.3 cm$^{-1}$) corresponding to the outer edge of
the A ring to the inner edge of the D ring. (high $E_p$'s) and (low $E_p$'s) are not in this range.

The comment columns contain the indices [$i$]'s used in Tables 6, 7, 8, 9 and Figs 12 and 13.

Other comments connect certain $E_p$'s with the A, B, C and D rings.

Bolded $E_p$'s with [$i$]'s associated with them are the key $E_p$'s in Fig.13.

The center of the B ring is near the peak in the PED of Saturn's protosatellite disk.
The inner B ring exists to the left of this peak in Fig. 13.
The outer B ring exists to the right of this peak in Fig. 13.

Table 9. Photon energies ($E_p$'s) and orbital radii ($r/R_s$) used for the construction of Figs. 13, 16, 17 and 18. The individual pairings of $E_p$'s and $r/R_s$'s are determined by first pairing $E_p(7,\infty)$ with the inner radius of the A ring and the inner and outer radius of the C ring. Then the other pairings automatically fall into place.

| Satellite or Ring Edge Name | [i][a] | $r/R_s$ [b] | $n_f, n_i$ [c] | $E_p(n_f, n_i)$ [c] (cm$^{-1}$) |
|---|---|---|---|---|
| Saturn's Equatorial radius |  | 1.000 |  |  |
| Inner Edge D ring | [29] | 1.110 | 6,∞ | 3048.3 |
| D68 ringlet | [28] | 1.122[d] | 5,9 | 3034.7 |
| D72 ringlet | [27] | 1.187[d] | 5,8 | 2674.8 |
| D73 ringlet | [26] | 1.216[d] | 4,5 | 2469.1 |
| Outer Edge D ring |  | 1.236 |  |  |
| Inner Edge C ring | [25] | 1.239 | 7,∞ | 2239.5 |
| Titan ringlet in Colombo Gap | [24] |  | 5,7 | 2149.9 |
| Titan ringlet in Colombo Gap | [24] |  | 6,11 | 2141.3 |
| Average Titan ringlet $E_p$'s | [24] | 1.292 |  | 2145.7 |
| Maxwell ringlet in Maxwell Gap | [23] |  | 6,11 | 2141.3 |
| Maxwell ringlet in Maxwell Gap | [23] |  | 5,7 | 2149.9 |
| Average Maxwell ringlet $E_p$'s | [23] | 1.452 |  | 2145.7 |
| Outer Edge C ring | [22] | 1.526 | 7,∞ | 2239.5 |
| Inner Edge B ring |  | 1.526 |  |  |
| Outer Edge B ring |  | 1.950 |  |  |
| Huygens ringlet in Huygens Gap | [21] | 1.955[e] | 6,13 | 2398.9 |
| Laplace ringlet in Laplace Gap | [20] | 1.992[e] | 6,12 | 2286.2 |
| Inner Edge A ring | [1] | 2.030 | 7,∞ | 2239.5 |
| Pan in Encke Gap | [2] |  | 5,7 | 2149.9 |
| Pan in Encke Gap | [2] |  | 6,11 | 2141.4 |
| Average Pan $E_p$'s | [2] | 2.217 |  | 2145.7 |
| Daphnis and Keeler Gap | [3] | 2.265 | 7,20 | 1965.2 |
| Outer Edge A ring |  | 2.270 |  |  |

[a] Indices in Tables 7, 8 and 9 & Figs. 12, 13 and 15.

[b] Orbital radii of satellites and rings in units of the equatorial radius of Saturn and from NASA (2021) except as otherwise noted. These orbital radii are transformed with Eq. (8) to give $(r/R_s)'$ values used in Fig. (18).

[c] The quantum numbers that define transitions in the hydrogen atom and photon energies associated with these transitions.

[d] Hedman et al. (2007b)

[e] French et al. (2020) their Fig. 2 and NASA (2022)

**Fig. 13. The Photon Energy Distribution in Saturn's protoplanetary disk, highlighting the A, B, C and D rings & the Cassini Division**

**Figure 13**

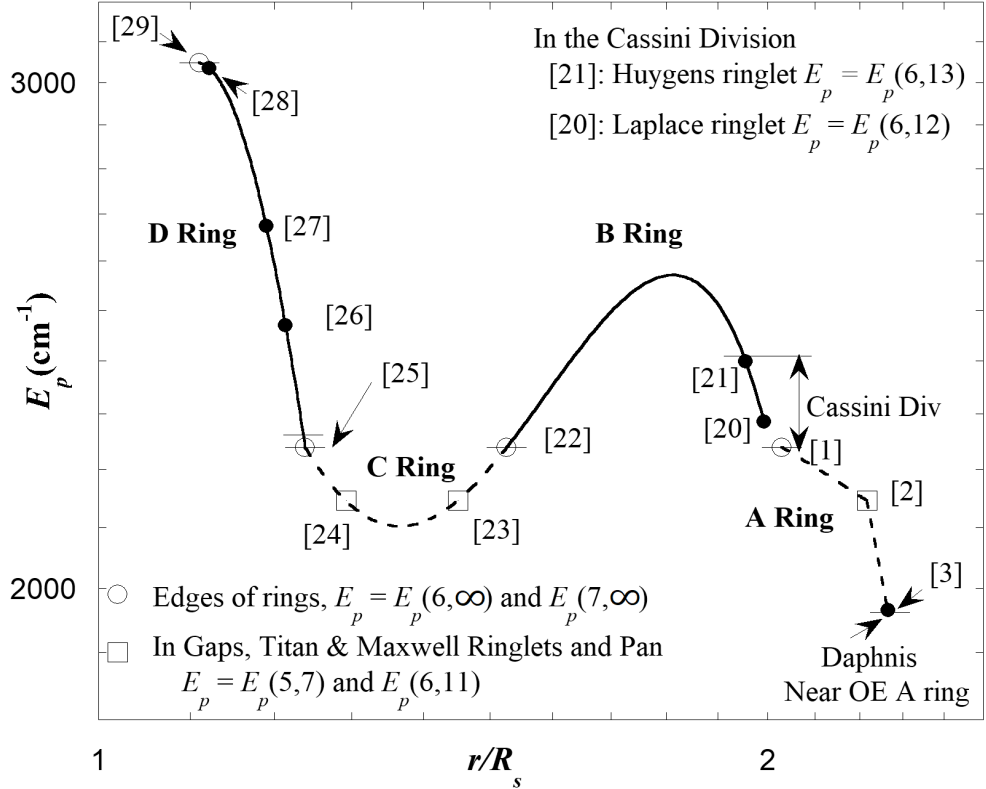

Furthermore, Table 8 reveals there are just three bolded $E_p$'s that fall within the D ring. These $E_p$'s are associated with the three ringlet radii. Figure 13 shows the resulting PED from index [22] to index [29] to be a smooth curve that defines a dip in the PED of Saturn's protosatellite disk.

*2.7.c. The B Ring, Cassini Division and Resonance Ring Widths*

In the present model the broad rings of Saturn are created by overlapping adjacent resonance rings. Outside the region of Saturn's broad rings, resonance rings do not overlap. Whether they overlap or not depends on their width. The widths of resonance rings within the wide rings undoubtedly depend on a few factors, e.g. the local temperature gradient, and the dust and gas densities in the rings. The variability in these factors could explain why the inner edge of the B ring extends inward all the way to the outer edge of the C ring while the outer edge does not extend outward all the way to the inner edge of the A ring, and the Cassini Division exists between the A and B rings. On average the optical depth of the Cassini Division is low (French et al. 2020), but it is not completely void of material and there are eight gaps (Planetary Society 2021) in the Cassini Division. Among these are the Laplace and Huygens Gaps each with a ringlet within its boundaries.

The optical depth of the two ringlets is large and the ringlets are narrow compared to the other features in the Cassini Division, as seen in the optical depth profile of the Cassini Division (French et al. 2020 Fig. 2 & Jerousek et al. 2020 Fig.1). Therefore, the Laplace and Huygens ringlets are the dominant features in the Cassini Division. In Table 8 there are just two $E_p$'s, $E_p(6,12)$ and $E_p(6,13)$, that have values corresponding to the Cassini Division and they are paired with the Laplace [20] and Huygens [21] ringlets in Table 9 and Figs. 13 and 14. The width of these ringlets are about 40 and 20 km respectively and the difference in their radii is about 2,200 km (French et al. 2020 Fig. 2).

**Fig. 14. The PED in the B ring and Cassini Division with the middles of resonance rings indicated as well as quantum numbers of corresponding $E_p$'s**

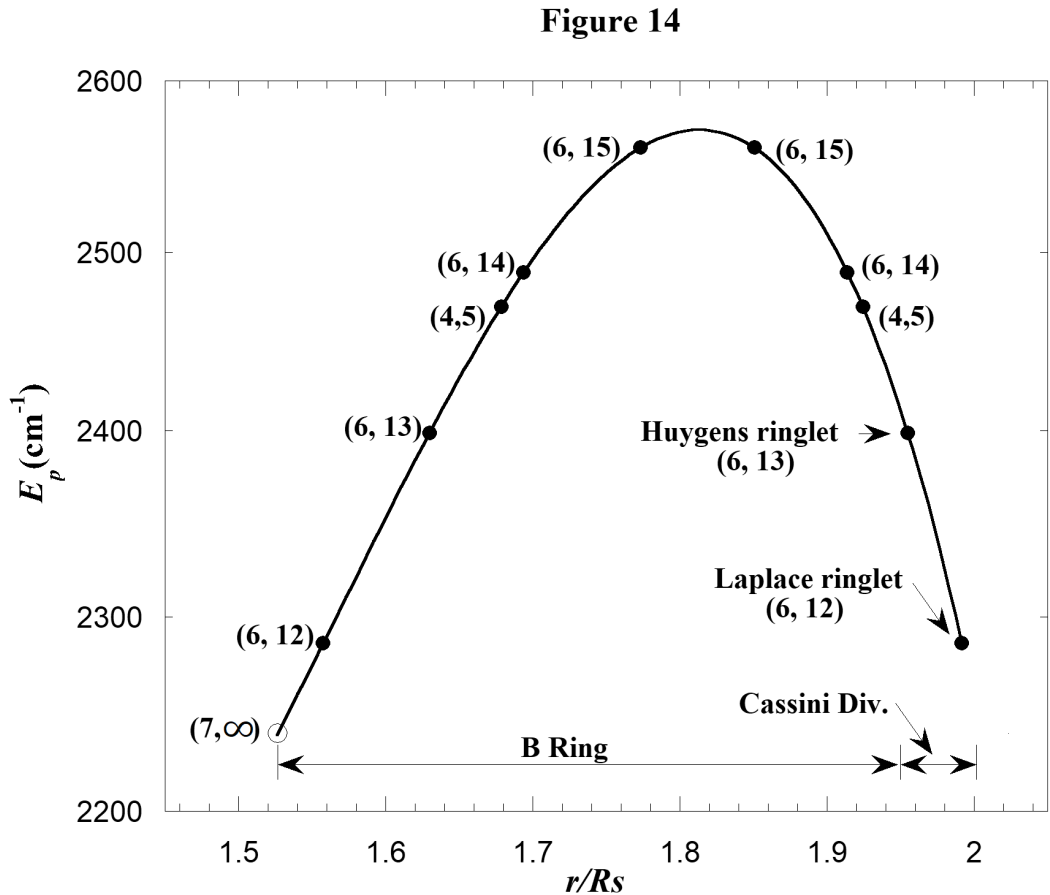

**Figure 14**

The peaked solid curve in Fig. 13 approximates the PED in the region of the B ring and Cassini Division. It is determined as described in Appendix 4. In this approximation the peak value in the PED is slightly larger than $E_p(6,15)$. Therefore $E_p(6,15)$ is the largest $E_p$ value contributing to the B ring as indicated in Table 8. Using Table 8 and Fig. 14, we can count the number of just barely overlapping resonance rings that create the B ring by counting the $E_p$'s that create resonance (Three of them are counted twice). Starting from the inner edge of the B ring and moving outward to the outer edge we have $E_p(6,12)$, $E_p(6,13)$, $E_p(4,5)$, $E_p(6,14)$, $E_p(6,15)$, $E_p(6,15)$, $E_p(6,14)$, $E_p(4,5)$. There are 8 overlapping resonance rings. Assuming they barely overlap, the average width of a resonance ring is approximately 1/8th the width of the B ring or approximately 3200 km. This is about 100 times wider than the Laplace and the Huygens ringlets in the Cassini Division where the optical depth is low.

In Table 8, [26] is paired with $E_p(4,5)$ and this resonance causes a ringlet in the D ring. We might expect photons with energy $E_p(4,5)$ to affect the B ring in the same way they affect the D ring. If this were the case, there would be two ringlets corresponding to the $E_p(4,5)$ resonance also observed in the B ring with one on each side of the PED peak. The lack of these ringlets is likely related to the larger density of material that presumably existed in the B ring over the D ring when resonance rings existed. Presently the optical depths of the B and D rings are 0.4-2.5 and $10^{-5}$ respectively (NASA 2021).

### 2.8. The Temperature Distribution in Saturn's Protosatellite Disk

Mousis, Gautier and Bockelee-Moran (2002) (in the future referred to as MGB) have theoretically derived a series of six TD's for the Saturnian subnebula (protosatellite) disk. Their TD's are shown in their Fig. 1. The first four of these are reproduced in Fig. 15. (The reproduction is achieved by analyzing coordinates of points on the MGB TD curves with the photo editor GIMP). We now investigate the relationship between the MGB TD's and the $E_p$'s and $r/R_s$ values listed in Table 7.

**Fig. 15. Reproduction of TD's in Saturn's protosatellite disk***

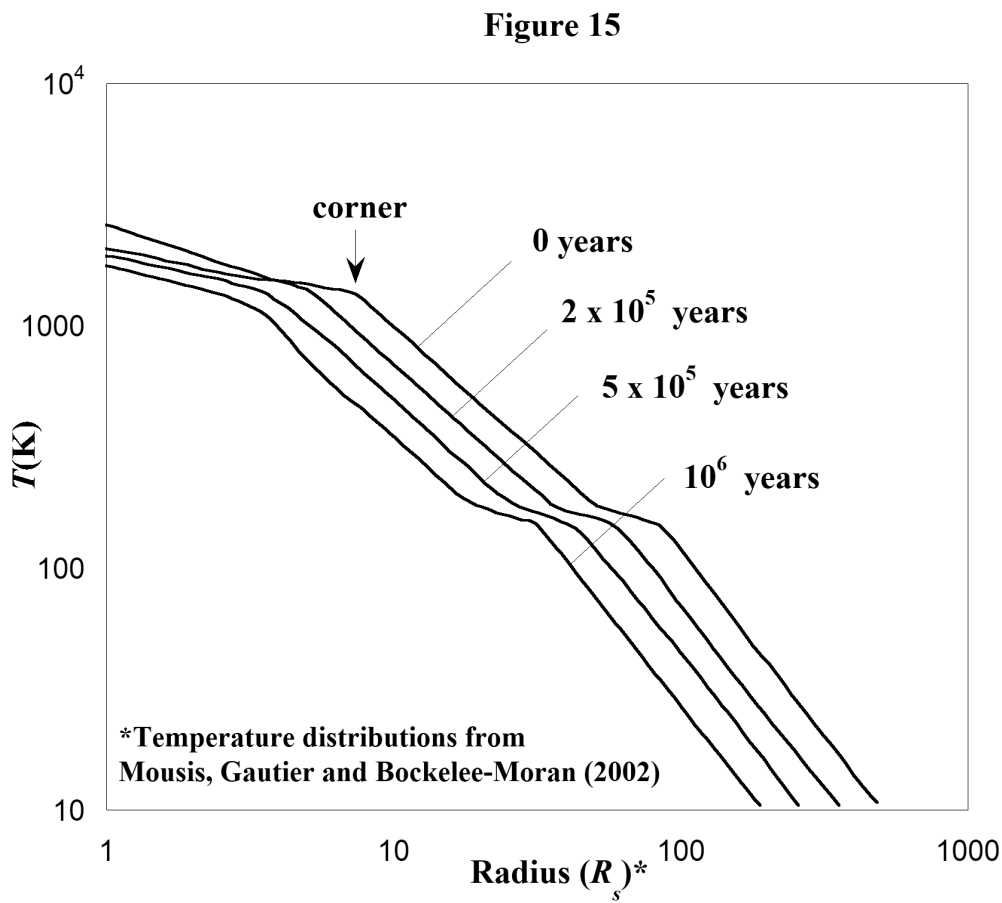

*Temperature distributions from Mousis, Gautier and Bockelee-Moran (2002)

As in section 2.3.b for Uranus, we assume Saturn's rings and regular satellites evolved from primordial rings that were born in the Saturnian protosatellite disk with each primordial ring associated with the local mid-plane temperature $T$ of the disk. Each of these $T$'s is related to one of the $E_p$ values previously discussed for the Saturnian disk. We again assume the relationship between $T$ and $E_p$ is linear and of the form

$$T = C'_1(E_p + C'_2), \qquad (7)$$

where $C'_1$ and $C'_2$ are constants to be empirically determined. To determine these constants we use Eq. (7) with the $r/R_s$ values and $E_p$'s in Table 7. A trial set of $C'_1$ and $C'_2$ is used to calculate a $T$ for each $E_p$ in Table 7. Then these $T$'s and their corresponding values of $r/R_s$ are used to construct a trial TD. This TD is compared to the TD curves in Fig. 15. After many iterations it is determined that no good fit can be found to any of the MGB TD's. Fig. 16 shows the fit to the MGB TD for which $t = 5 \times 10^5$ years. The fit is not good but the shape of the trial TD (the collection of data points) gives a hint on how to proceed. A corner in the trial TD is identified in Fig. 16. And notice the MGB TD for which $t = 0$ in Fig. 15 also has a corner. Perhaps these corners can be made to overlap and hopefully this will produce a good fit of the Table 7 data to the MGB TD for which t = 0 years. A linear scaling of the $r/R_s$ values in Table 7 to larger values, call them $(r/R_s)'$, is assumed.

**Fig. 16. Fitting the Saturnian disk TD to the MGB TD for which $t = 5 \times 10$ years. The fit is not as good as the one in Fig. 17.**

**Figure 16**

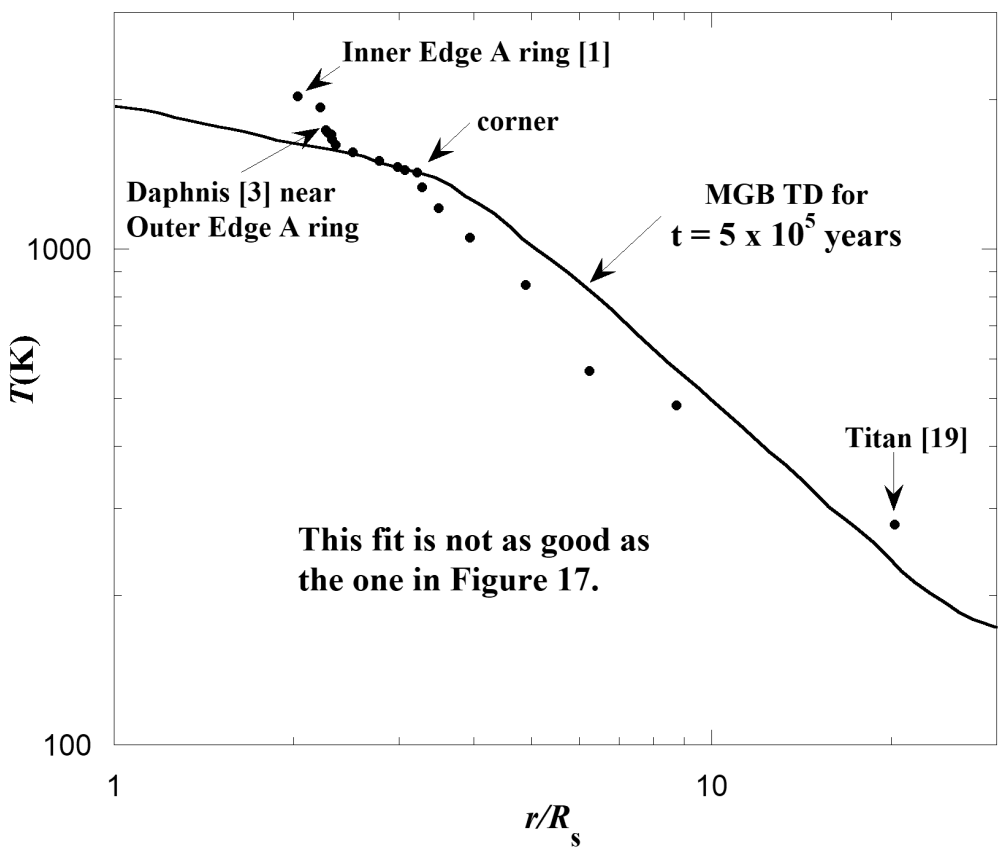

$$(r/R_s)' = -a + (a+1)(r/R_s), \qquad (8)$$

where $a$ is a third adjustable constant along with $C'_1$ and $C'_2$. Perhaps, if the MGB TD's in Fig. 15 were to be linearly compressed by reasonably altering parameters used to calculate these TD's, it would not be necessary to transform orbital radii by means of Eq. (8). This form for Eq. (8) ensures that orbital radii of satellites and rings that are close to Saturn's surface are not altered very much. (Note: When $(r/R_s) = 1$ so does $(r/R_s)'$). Also the larger the orbital radius the larger the alteration. Subsequently the $(r/R_s)'$ values, not the $r/R_s$ values, are used in the fitting process.

### 2.8.a. Fitting the MGB TD for which t = 0 years

The data from Table 7 is used to fit the MGB t = 0 year TD by adjusting the constant $a$ in Eq. (8) along with $C'_1$ and $C'_2$. The best fit parameters are $a = 2.20$, $C_1' = 1.080$ K·cm and $C_2' = -462$ cm$^{-1}$. The best fit to the t = 0 year TD is in Fig. 17.

Fig. 17. The best fit to any MGB TD is achieved by transforming satellite $r/R_s$ values to $(r/R_s)'$ values. This MGB TD corresponds t = 0 years.

Figure 17

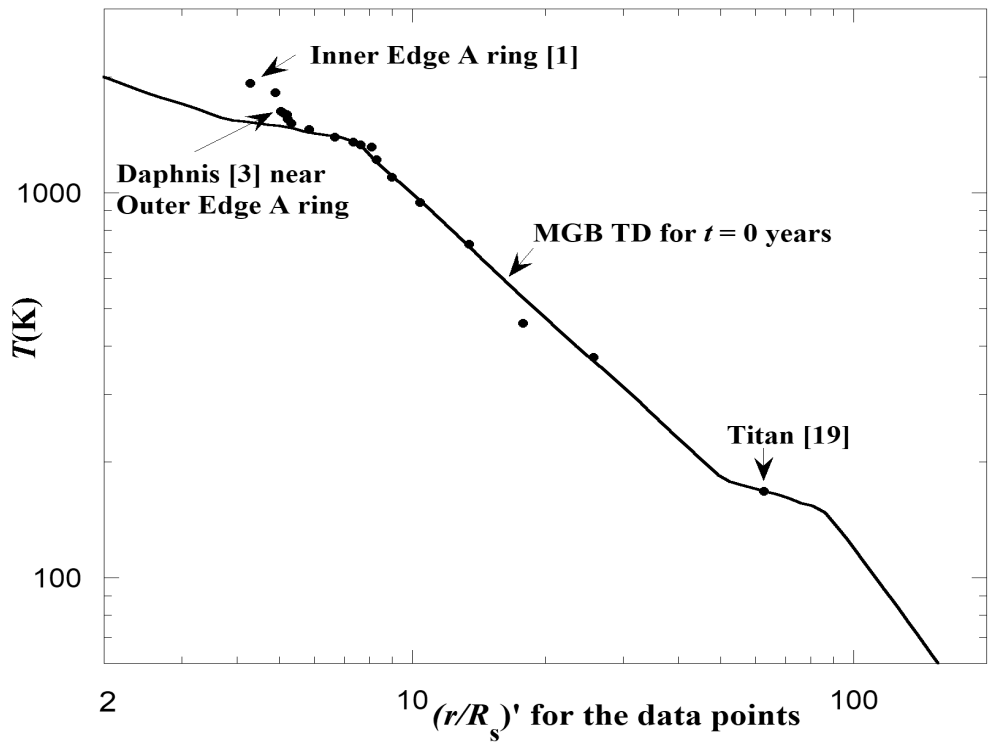

Eq. (7) and the best fit parameters are used with Tables 7 and 9 to determine the complete Saturnian disk TD seen in Fig. 18. Also shown there is the MGB TD for which $t = 0$ years. The complete TD found in the present investigation has a peak just as in the Uranian disk TD. Note the present model TD diverges from the MGB TD just outside the outer edge of Saturn's A ring which is positioned on the right side of the peak. This divergence is similar to the one in the Uranian and Neptunian disk TD's in Fig. 10. Additionally, just up from the point of divergence there is a small shoulder in the side of the TD's peak. This is reminiscent of the small shoulder in a similar spot in Fig. 10.

**Fig. 18. The overlap of the complete Saturnian TD from the present investigation with the MGB TD for which $t = 0$ years.**

### Figure 18

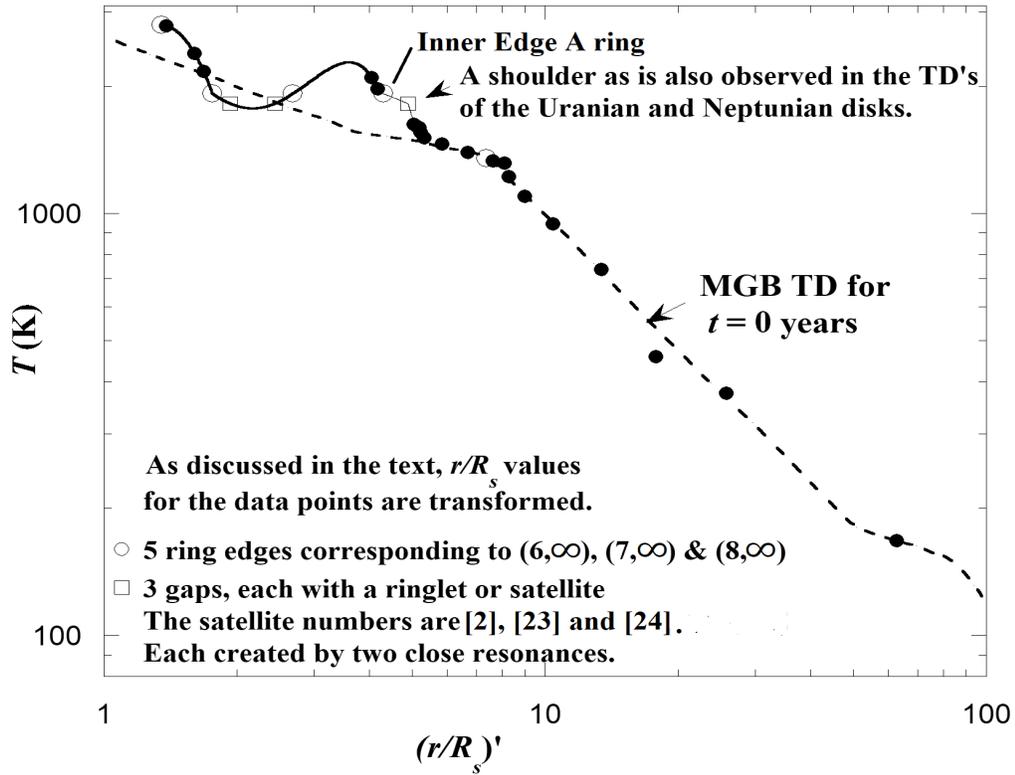

With the determined values of $C_1' = 1.080$ K·cm and $C_2' = -462$ cm$^{-1}$ substituted into Eq. 7 we have

$$T = 1.08\text{K·cm} (E_p - 462\text{cm}^{-1}), \qquad (9)$$

In section 2.6 we compared the empirically determined $T(E_p)$ (Eq. (6a)) for the Uranian satellite investigation to the theoretical $T(E_p)$ (Eq. (5)). These equations are

$$T = (hc/k)(E_p - \Delta) \qquad (5)$$

and
$$T = 1.609(hc/k)(E_p - 3720 \text{ cm}^{-1}). \quad \text{Uranian satellites} \qquad (6a)$$

Similarly we now compare the empirically determined $T(E_p)$ (Eq. (9)) for the Saturnian satellite investigation to Eq. (5) by writing Eq. (9) in terms of the factor $hc/k$. The result is

$$T = 0.751(hc/k)(E_p - 462\text{cm}^{-1}), \quad \text{Saturnian satellites} \quad (9a)$$

where $0.751(hc/k) = C_1' = 1.080$ K·cm.

The leading constants (1.609 and 0.751) on the right sides of Eqs. (6a) and (9a) are both in reasonable agreement with the leading constant (1) on the right side of the theoretical relationship Eq. (5). However the fact that the constants (1.609 and 0.751) differ by nearly a factor of 2 is interesting. As mentioned before, Appendix 3 contains the derivation of Eq. (5) for the case of $A$ and $B$ colliding and associating to form $AB$. It also includes a similar derivation for a relationship between $T$ and $E_p$ that in addition involves third body assisted stimulated radiative molecular association (3rdBA SRMA). The proposed reaction is discussed in Appendix 3. The result is

$$T = (hc/2k)(E_p - \Delta). \quad (10)$$

Note the additional factor of 2 in the denominator of Eq. (10) compared to Eq. (5) above.

We compare Eq. (9) to Eq. (10) by rewriting Eq. (9) in terms of the factor $hc/2k$. The result is

$$T = 1.502(hc/2k)(E_p - 462\text{cm}^{-1}), \quad \text{Saturnian satellite} \quad (9b)$$

where $1.502(hc/2k) = C_1' = 1.080$ K·cm.

The leading constant 1.502 in Eq. (9b) is in good agreement with the leading constant of 1.609 in Eq. (6a). The agreement suggests that 3rdBA SRMA could be the key reaction that triggered satellite evolution in the Saturnian disk. The approximations made in Appendix 3 are similar for both cases presented there. Therefore if they were relaxed in both, possibly the leading constants in Eqs. (6a) and (9b) would still be in agreement.

## 3. FURTHER DISCUSSION AND CONCLUSIONS

If the collection of matter that leads to satellites in protoplanetary disks were mainly due to the force of gravity we would not expect the plots of Uranian satellite orbital radii vs. Jovian satellite orbital radii (Fig. 1) and Uranian satellite orbital radii vs. Neptunian satellite orbital radii (Fig. 2) to be so well fitted by straight lines. Also this investigation indicates that satellite migration is not a large factor in the evolution of planetary systems unless it is a uniform effect without satellites crossing orbits.

The empirically determined value for $\Delta = 3720$ cm$^{-1}$ in Eq. (6a) may be helpful in the identification of the reactants $A$ and $B$. The energy $\Delta$ is associated with a transition the molecule $AB$ undergoes during the molecule's association. Black and van Dishoek (1987) list transition energies and relative spectral intensities in $H_2$. The (1,0)O(3) transition has an energy 3568 cm$^{-1}$ and spectral intensity considerably higher than all other transitions near it in the spectrum. The energies (3720 and 3568 cm$^{-1}$) are in reasonable agreement. Similarly $\Delta = 462$ cm$^{-1}$ in Eq. (9b) is in reasonable agreement with the transition energy 354 cm$^{-1}$, associated with the (0,0)S(0) transition in $H_2$, which has the largest spectral intensity in the $H_2$ spectrum (Black and van Dishoek 1987). Furthermore, Latter and Black (1990) study the formation of $H_2$ by radiative association. The reaction mechanism in their study is similar to SRMA except it is without stimulation. They show their mechanism requires a stabilizing photon be emitted by the $H_2$ during its formation. In the present model outlined in section 2.6 the stabilizing photon's energy is $E_p = |\Delta K| + \Delta$, where $\Delta$ is possibly the energy of the (1,0)O(3) or (0,0)S(0) transition in $H_2$.

The Colombo, Maxwell and Encke Gaps and their ringlets are produced by two closely spaced resonances associated with $E_p(5,7)$ and $E_p(6,11)$. The two together seemingly caused resonances that were strong enough to clear the gap regions and produce ringlets. Without this fortunate situation and without the well defined positions of ring edges associated the series limit $E_p(7,\infty)$, it would be impossible to determine the shape of the Saturnian Disk TD in the region of the C ring.

The present model predicts resonance rings are created early in the evolution of the Uranian and Saturnian protosatellite disks. Furthermore it seems that the matter that forms satellites is efficiently collected in these rings. If these are true, the accretion time scale is most certainly shortened from what it would be if no resonance rings existed.

The smoothness of the Uranian disk PED in Figs. 5a and 5b supports the model presented in this paper. The only point on the PED noticeably displaced corresponds to ring η. As pointed out in subsection 2.2.b this displacement could be due to gravitational resonant interaction with the satellite Cressida (Chancia et al. 2017). Also supporting the present model is the good quality of the fits of the three TD's determined in this investigation for the Uranian, Jovian and Neptunian disks to the Mousis (2004) TD (see Fig. 10) as well as the Saturnian disk TD fit to the Mousis, Gautier and Bockelee-Moran (2002) TD (see Fig. 17).

Results of the present investigation indicate rings of Saturn developed at the same time all its regular satellites developed. This is in conflict with findings from the Cassini spacecraft. Cassini measurements reveal a lower than expected total ring mass which in turn suggests the rings are only 10-100million years old (Iess et al. 2019).

Interestingly this investigation indicates the possibility that the quantum nature of matter has put its stamp not only on atomic but also on planetary systems.

## ACKNOWLEDGMENTS

I would like to acknowledge members of the Allegheny College Physics Department for supporting my studies in this area for many years, especially Professors Richard Brown and James (Jamie) Lombardi Jr. Many thanks to Jamie for his editorial assistance with this paper, especially for suggesting that the factors (hc/k) and (hc/2k) be included in Eqs. (6a) and (9b) for easier comparison of various equations and for suggestions concerning the discussion of the derivations in Appendix 3. Also, many thanks to Jerry Staub for reading and commenting on various versions of this present paper and to Jim Fitch for reviving my desire to continue working in this area. I also acknowledge the tremendous amount of experimental and theoretical work carried out by others that is essential for this investigation.

# APPENDIX 1
*Mousis Temperature Distributions*

*Section A1.a. Reproduction of Mousis TD's in Figure 6*

Figure 6 includes reproductions of three TD's from Mousis (2003). These are the three TD's that are labeled $t = 1 \times 10^4$, $2 \times 10^4$ and $5 \times 10^4$ years. The graphic editor GIMP is used to determine the coordinates of critical points in Figure 1 of Mousis (2003). Each of these TD's is represented mostly by straight line segments when plotted using logarithmic scales. The only exception is the curved, very top portion of the $t = 1 \times 10^4$ TD which is not reproduced in Figure 6 of the present investigation. Each of the straight line segments in Figure 6 is of the form $T = \beta r^n$, where $r$ is the radial coordinate in the disk and $\beta$ and $n$ are constants that determine the position and slope of the straight line segments. The various values of $\beta$ and $n$ are used to reproduce the Mousis TD's in Figure 6.

*Section A1.b. Construction of Dashed TD's in Figure 6*

Similarities among the three Mousis TD's make it possible to construct the two dashed TD's in Fig.6.

1. The top segment of each Mousis TD connects to the top of its middle segment at ***approximately*** the same temperature. And the bottom segment of each TD connects to the bottom of its middle segment at ***approximately*** the same temperature. Call these temperatures $T_1$ and $T_1'$ in the TD for which $t = 1 \times 10^4$ years.
2. Each segment in a TD is nearly parallel to a corresponding segment in the other two TD's. I.e. they have nearly the same $n$. Call these $n$ values, $n_1$, $n_2$ and $n_3$ in the TD for which $t = 1 \times 10^4$ years.

The dashed TD's in Fig.5 are just two in a series of trial TD's constructed for the present investigation. Just one of these TD's (for which $t = t_0 = 8800$ years) is fitted well by the fitting process described in subsection 2.3.c. Because the dashed TD's in Figure 5 are close to the Mousis $t = 1 \times 10^4$ year TD, the approximate similarities mentioned above can be taken to be nearly exact and each trial TD is determined as follows.

1. Temperatures where a trial TD's segments connect are $T_1$ and $T_1'$, the temperatures mentioned in similarity 1 mentioned above.
2. Its three segments have $n$ values, $n_1$, $n_2$ and $n_3$ just as in similarity 2 mentioned above.
3. The value of $\beta$ for the middle segment of a trial TD is taken to be a few percent larger than what it is in the Mousis TD for which $t = 1 \times 10^4$ years. Appendix 2 explains how $t$ for a trial TD is determined from this new $\beta$.
4. The two radial coordinates $r_1$ and $r_1'$, for the points where segments connect, can be found by applying the relationship $T = \beta r^n$ to each end of the middle segment where $T$ is either $T_1$ or $T_1'$.
5. Repeated use of the equation $T = \beta r^n$, determines the $\beta$ values for the rest of the segments.
6. Knowing $\beta$ and $n$ for each segment determines the trial TD.

## APPENDIX 2

*Finding the Time When the Uranian Satellites Begin Their Evolution*

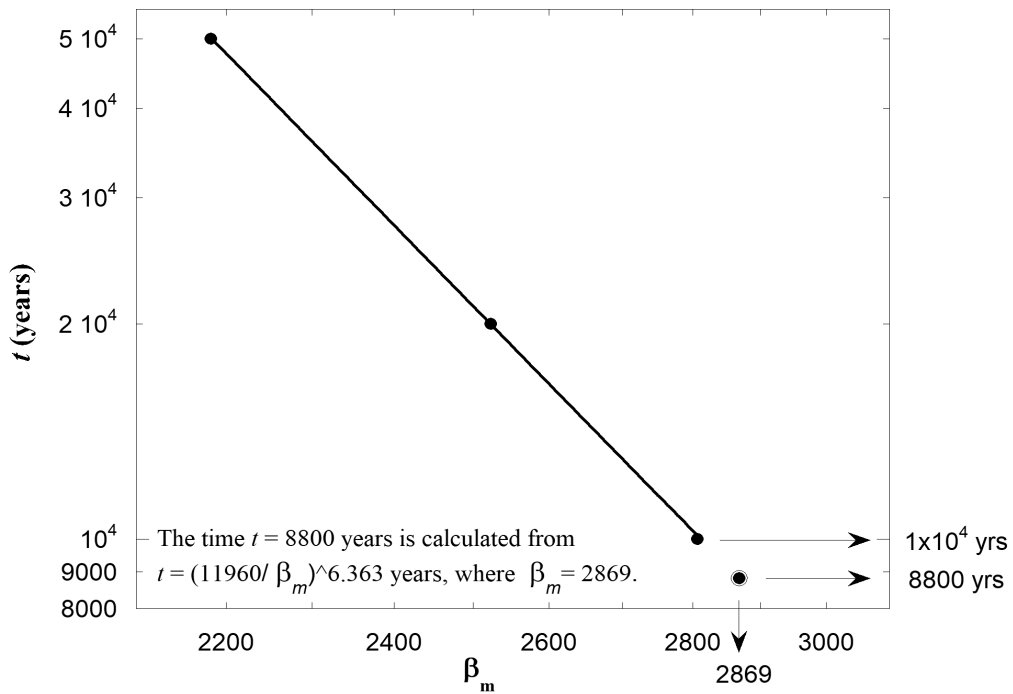

**Figure A2**
**The graph of log($t$) vs log($\beta_m$) is linear.**

The time $t = 8800$ years is calculated from $t = (11960/\beta_m)^{6.363}$ years, where $\beta_m = 2869$.

The three TD's in Figure 6 associated with $t = 1 \times 10^4$, $2 \times 10^4$, and $5 \times 10^4$ years are all reconstructed from Figure 1 in Mousis (2003). Each segment of these TD's is characterized by an equation of the form $T = \beta r^n$, as discussed in Appendix 1. Figure A2 utilizes the values of $\beta$ for the three middle segments for the reconstructed Mousis TD's: we are calling them $\beta_m$. The values of $\beta_m$ are measured, as described in Appendix 1, to be 2810, 2520 and 2180 for times $t = 1 \times 10^4$, $2 \times 10^4$, and $5 \times 10^4$ years respectively. Because the graph of log($t$) vs log($\beta_m$) is linear, the relationship between $t$ and $\beta_m$ has the form

$$t = \text{constant} \times \beta_m^{-P}$$

and the best fit to the data is $t = (11960/\beta_m)^{6.363}$. The well fitted TD in Fig.8 has $\beta_m = 2870$. We extrapolate the graph in Figure A2 to earlier time by substituting this value of $\beta_m$ into the above equation. The result for $t$ is $t_0 = 8800$ years, the time when the Uranian satellites begin their formation. This is also the $t$ value for one of the dashed TD's in Fig. 6. For the other dashed TD in Fig 6, $\beta_m$ is arbitrarily taken to be 6% larger than the $\beta_m$ in the Mousis TD for which $t = 1 \times 10^4$ years. The $t$ value for the other dashed TD is found in the same way to be 6900 years.

## APPENDIX 3

*The Relationship Between T and $E_p$*

*Section A3.a. Case 1: Protosatellite Disks of Uranus, Jupiter and Neptune*

We use the model presented in section 2.6 to derive $T$ as a function of $E_p$. The following relationship, Eq. (4) in the main text, is an expression of the SRMA that is key to the present discussion,

$$A + B + h\nu \rightarrow AB + 2h\nu. \quad (4)$$

In this reaction $A$ and $B$, with momenta $p_A$ and $p_B$, collide to form $AB$ with the momentum $p_{AB}$. A more accurate calculation to determine $T(E_p)$ would include the realistic effect of a range of possible angles between $p_A$ and $p_B$ before the collision. However a likely angle between $p_A$ and $p_B$ is $90°$. For the sake of simplifying the calculation, we calculate $T(E_p)$ for the single case of $p_A$ and $p_B$ being perpendicular

$$p_{AB}^2 = p_A^2 + p_B^2. \quad (A3.1)$$

Also the momentum of $p_{AB}$ is about 30,000 times greater than momentum of the photon created during the collision. Therefore there is no need to account for the photon's momentum in Eq. (A3.1). The kinetic energies of $A$, $B$ and $AB$ are $K_A = \frac{1}{2}p_A^2/m_A$, $K_B = \frac{1}{2}p_B^2/m_B$, and $K_{AB} = \frac{1}{2}p_{AB}^2/m_{AB}$, where $m_A$, $m_B$, and $m_{AB}$ are the respective masses and $m_{AB}$ has a value that is extremely close to $m_A + m_B$. With these expressions, Eq. (A3.1) becomes

$$m_{AB} K_{AB} = m_A K_A + m_B K_B \quad (A3.2)$$

There is likely a range of kinetic energies associated with the $A$'s and $B$'s that participate in SRMA resonance. However we simplify our calculation by defining $K_{mp}$ as the most probable kinetic energy of the $A$'s and $B$'s that participate. The most probable speed in a Maxwell-Boltzmann distribution is $v_{mp} = (2kT/m)^{1/2}$, where $m$ is the particle mass. Particles with this speed have a kinetic energy of $kT$ and we take $K_A = K_B = K_{mp} = kT$.

All of this together yields $\quad K_{AB} = K_{mp}(m_A + m_B)/m_{AB} = K_{mp},\ $ (final kinetic energy) (A3.3)

where $(m_A + m_B)/m_{AB}$ is extremely close to unity.

Also $\quad\quad\quad\quad\quad\quad K_A + K_B = 2K_{mp} \quad$ (initial kinetic energy) $\quad\quad$ (A3.4)

and the change in kinetic energy of the system during the association that creates $AB$ is

$$\Delta K = (K_{AB} - (K_A + K_B)) = -K_{mp} = -kT. \quad (A3.5)$$

By conservation of energy, the energy of the created photon ($h\nu$) is

$$E_p = |\Delta K| + \Delta, \quad (A3.6)$$

where the energy $\Delta$ accounts for a transition that may occur in the molecule during its association. From Eqs. (A3.5) and (A3.6)

$$T = (1/k)(E_p - \Delta). \tag{A3.7}$$

In the analysis $E_p$'s are in the units of wave numbers. To account for this we multiply the right side of Eq. (A3.7) by $hc$ where $h$ is Planck's constant and $c$ is the speed of light. Also the value of $(hc/k)$ is 0.014388 K·m = 1.4388 K·cm. Therefore we have

$$T = (hc/k)(E_p - \Delta). \tag{A3.8}$$

This is also Eq. (5) in the main text.

### A3.b. Case 2: Third Body Assisted Stimulated Radiative Molecular Association as the Possible Key Reaction that Triggered Satellite Evolution in the Saturnian disk

For Saturn's disk, Eq (9) in section 2.8 is determined by adjusting the parameters $a$, $C_1'$ and $C_2'$ so that $T$'s calculated from Eq. (7) fit the MGB (2002) TD for which t = 0 years. The result is

$$T = (1.080 \text{ K·cm})(E_p - 462 \text{ cm}^{-1}) \quad \text{(from fit)}. \tag{9}$$

Because the constant 1.080 K·cm in Eq. (A3.8) is about one half the constant 2.315 K·cm in Eq. (6) for Uranus's disk, we expect the key SRMA reaction for Saturn's disk to involve a different number of atoms or molecules. To include a third body in the reaction we consider the mechanism of third body assisted association (Fraser, McCoustra, and Williams 2002 & Turk et. al. 2011) characterized by

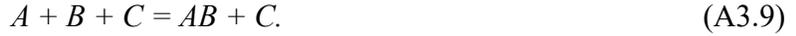
$$A + B + C = AB + C. \tag{A3.9}$$

In this reaction $C$ assists in the association of $A$ and $B$ to form $AB$. We propose the following reaction which is a combination of Eqs. (4) and (A3.9).

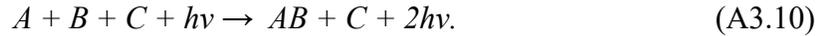
$$A + B + C + h\nu \rightarrow AB + C + 2h\nu. \tag{A3.10}$$

In Eq. (A3.10) both $C$ and $h\nu$ assist the radiative molecular association. Therefore Eq. (A3.10) characterizes third body assisted stimulated radiative molecular association. This is actually a 4-body interaction because of the requirement that a photon also must participate in the collision. Generally speaking 4-body interaction are less likely than 3-body interactions. However in this case the photon on the left side of Eq. (A3.10) is a readily available resonant photon with exactly the correct direction of motion to participate in a collision which in turn further enhances resonance. This should make the 4-body interaction more likely than it would otherwise be.

We now derive a formula that is close to the empirical Eq. (9) found for Saturn's disk. We make the simplifying approximation that $C$ moves off in the same direction and at the same speed as does $AB$ after the collision. But also physically speaking, it seems likely for $C$ to travel with $AB$ in order for $C$ to assist in the association of $A$ and $B$ to make $AB$. So the magnitude of the final momentum of the system is equal to the sum of the magnitudes of the momenta of $AB$ and $C$. In addition to formulas we used in Case 1 above, we have $K_{ABC} = \frac{1}{2}p_{ABC}^2/m_{ABC}$, where $K_{ABC}$ is the sum of the kinetic energies of $AB$ and $C$, (i.e. the final kinetic energy of the system) and $p_{ABC}$ is the magnitude of the system's momentum. Also, $m_{ABC} = m_{AB} + m_C$, where $m_C$ is the mass of $C$ and $K_C$ is its kinetic energy before the collision. The vectors $\boldsymbol{p}_A$, $\boldsymbol{p}_B$ and $\boldsymbol{p}_C$ are the momenta of $A$, $B$ and $C$ before they collide. Again, while realizing there is likely a range of angles between these vectors, we simplify the calculation by taking these vectors to be perpendicular for all resonant reactions. We therefore consider *A*, *B* and *C* approaching each other as if each one is moving along one of the axes in a three-dimensional Cartesian coordinate system. Therefore

$$(p_{ABC})^2 = p_A^2 + p_B^2 + p_C^2, \qquad (A3.11)$$

and
$$(m_{AB} + m_C) K_{ABC} = m_A K_A + m_B K_B + m_C K_C \qquad (A3.12)$$

Similarly to Case 1 we assume $K_A = K_B = K_C = K_{mp}$ and $(m_A + m_B + m_C)/(m_{AB} + m_C) = 1$.

Therefore, $\quad K_{ABC} = K_{mp}(m_A + m_B + m_C)/(m_{AB} + m_C) = K_{mp}$. (final kinetic energy) $\quad$ (A3.13)

Also $\qquad\qquad\qquad K_A + K_B + K_C = 3K_{mp} \qquad$ (initial kinetic energy) $\qquad$ (A3.14)

And $\qquad \Delta K = (K_{ABC} - (K_A + K_B + K_C)) = -2K_{mp} = -2kT \;\; (\text{where } K_{mp} = kT). \quad$ (A3.15)

As in Case 1 $\qquad\qquad\qquad E_p = |\Delta K| + \Delta \qquad\qquad\qquad$ (A3.16)

And this time $\qquad\qquad\qquad T = (hc/2k)(E_p - \Delta) \qquad\qquad\qquad$ (A3.17)

Note the 2 in the denominator of the first factor on the right side of Eq. (A3.10). This 2 does not appear in Eq. (A3.8)

## Appendix 4

*The Construction of the Approximate Curve Used for the*
*B Ring Portion of the PED in Figs. 13 and 14*

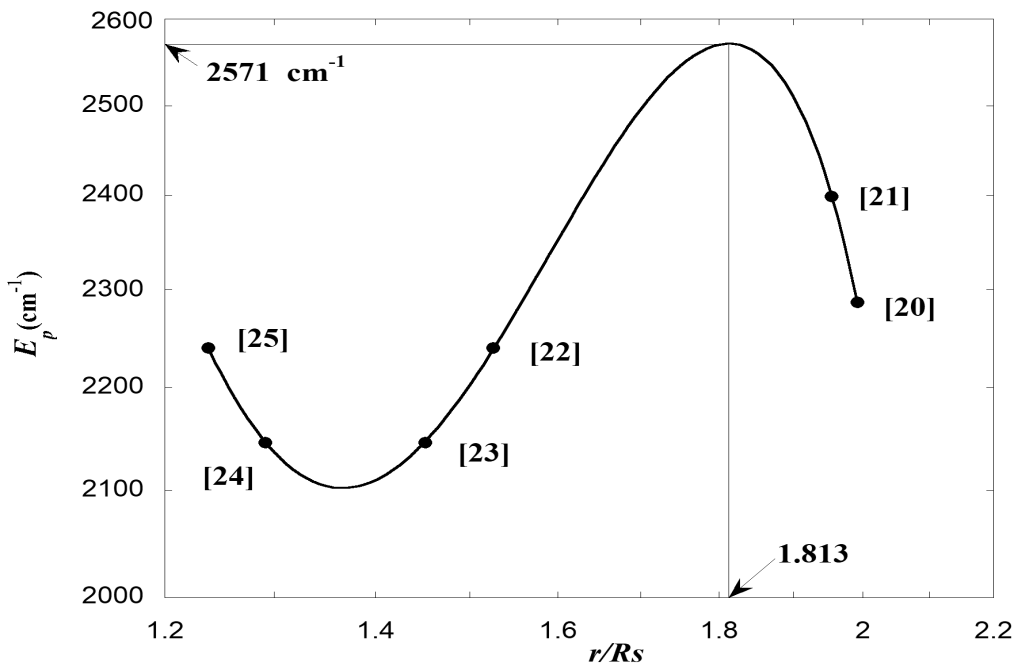

The present investigation does not indicate that any well defined resonance rings contribute to the creation of the B ring portion of the PED associated with Saturn's protoplanetary disk. Subsection 2.7.c indicates the B ring evolved from a number of wide resonance rings. Therefore it is necessary to make a reasonable approximation for the shape and magnitude of the PED in this region. This is achieved by (1) assuming the PED's value and slope are continuous across the boundary between the C and B rings and (2) assuming the PED passes through the two key points associated with the Huygens and Laplace ringlets. The four points that are associated with the C ring and the two points associated with the Huygens and Laplace ringlets are all fitted with one continuous function in the form of a cubic polynomial. This curve is used for the curve of the PED in the figure above and to fit the C ring and the Cassini Division as well as approximate the B ring in Figs. 13 and 14 in the main text.

An equation for the cubic polynomial in the figure is
$$E_p = 2571 - 7060(r/R_s - 1.8130)^2 - 10540(r/R_s - 1.8130)^3.$$

The curve has a maximum value of $E_p = 2571$ cm$^{-1}$ at $r/R_s = 1.8130$ and a minimum value of $E_p = 2102$ cm$^{-1}$ at $r/R_s = 1.3666$.

Italicized $r/R_s$ and $E_p$ data are used to make the fitted points in the graph.

| Satellite or Ring Edge Name | [i][a] | $r/R_s$[b] | $n_f, n_i$[c] | $E_p(n_f, n_i)$[c] (cm$^{-1}$) |
|---|---|---|---|---|
| *Inner Edge C ring* | [25] | *1.239* | *7,∞* | *2239.5* |
| Titan ringlet in Colombo Gap | [24] | | 5,7 | 2149.9 |
| Titan ringlet in Colombo Gap | [24] | | 6,11 | 2141.3 |
| *Average Titan ringlet $E_p$'s* | *[24]* | *1.292* | | *2145.7* |
| Maxwell ringlet in Maxwell Gap | [23] | | 6,11 | 2141.3 |
| Maxwell ringlet in Maxwell Gap | [23] | | 5,7 | 2149.9 |
| *Average Maxwell ringlet $E_p$'s* | *[23]* | *1.452* | | *2145.7* |
| *Outer Edge C ring* | *[22]* | *1.526* | *7,∞* | *2239.5* |
| Inner Edge B ring | | 1.526 | | |
| Outer Edge B ring | | 1.950 | | |
| *Huygens ringlet in Huygens Gap* | *[21]* | *1.955*[d] | *6,13* | *2398.9* |
| *Laplace ringlet in Laplace Gap* | *[20]* | *1.992*[d] | *6,12* | *2286.2* |

[a] Indices in Tables 8 and 9 and Fig. 14
[b] Orbital radii of satellites and ringlets in units of the equatorial radius of Saturn and are from NASA (2021) except as otherwise noted.
[c] The quantum numbers that define transitions in the hydrogen atom and photon energies associated with these transitions.
[d] French et al. (2020) Fig. 2 and NASA (2022)